\documentstyle[preprint,revtex,eqsecnum]{aps}
\begin{document}
\preprint{SISSA 217/92/EP}
\preprint{November, 1992}
\vskip 5.5truecm
\begin{title}
TOPOLOGICAL TWIST IN FOUR DIMENSIONS,\\
R-DUALITY AND HYPERINSTANTONS\\
\end{title}
\author{Damiano Anselmi and Pietro Fr\`e}
\begin{instit}
SISSA-International School for Advanced Studies\\
via Beirut 2, I-34100 Trieste, Italy\\
and I.N.F.N.- Sezione di Trieste, Trieste, Italy
\end{instit}
\eject
\begin{abstract}
ABSTRACT.
In this paper we continue the programme of topologically twisting
N=2 theories in D=4, focusing on the
coupling  of vector multiplets to  N=2 supergravity.
We show that in the minimal case, namely when the special geometry
prepotential $F(X)$ is a quadratic polynomial,
the theory  has a so far unknown
on shell $U(1)$ symmetry, that we name R-duality.
R-duality is a generalization of the chiral-dual on shell symmetry
of N=2 pure supergravity and of the R-symmetry of
N=2 super Yang-Mills theory. Thanks to this,
the theory can be topologically
twisted and topologically shifted, precisely as pure N=2  supergravity,
to yield a natural coupling of
topological gravity to topological Yang-Mills theory.
The gauge-fixing
condition that emerges from the twisting is the self-duality condition
on the gauge field-stength and on the spin connection. Hence our theory
reduces to intersection theory in the moduli-space of gauge instantons living
in gravitational instanton backgrounds. We remark that, for deep
properties of the parent N=2 theory, the topological
Yang-Mills theory we  obtain  by taking the flat space limit of our gravity
coupled Lagrangian is different from the Donaldson theory constructed by
Witten. Whether this difference is substantial and what
its geometrical implications may be is yet to be seen.

We also discuss the topological twist of the hypermultiplets leading to
topological quaternionic $\sigma$-models.
The instantons of these models, named by us hyperinstantons, correspond to
a notion of triholomorphic mappings discussed in the paper.

In all cases
the new ghost number is the sum of the old ghost number plus the R-duality
charge. The observables described by the theory are briefly
discussed.

In conclusion, the topological twist of the complete N=2 theory defines
intersection theory in the moduli space of gauge instantons plus gravitational
instantons plus hyperinstantons. This is possibly a new subject for
further mathematical investigation.
\end{abstract}

\section{Introduction}
\label{intro}

Topological Field Theories have stirred a lot of interest, both in
two and in four dimensions \cite{suit1}. Their general feature is that of
recasting intersection theory in the moduli-space of some suitable
geometrical structure into the language of standard quantum
field-theory, specifically into the framework of the path-integral.
Indeed the point-independent correlation functions of these peculiar
field-theories represent intersection integrals of cohomology classes in
the given moduli-space.

 Hystorically, the first topological field-theory
that has been introduced, is the topological version of 4D Yang-Mills
theory \cite{witten}, sometimes named Donaldson theory. It deals with the
moduli-spaces of Yang-Mills instantons and its correlation functions
describe Donaldson invariants \cite{donaldson}.
A lot of attention has also been
devoted  to topological sigma models in two dimensions
\cite{witten2}. In this case
one probes the moduli-space of holomorphic mappings from the world-sheet
to a complex target space. Theories that have a close relation
with topological sigma-models are the topological versions of N=2
Landau-Ginzburg models \cite{suitA}.
They have provided an interesting arena for
the study of the moduli-spaces associated with Calabi-Yau manifolds
\cite{suitB},
a topic  of primary interest in connection with the effective
Lagrangians of superstring models.
In a different, but closely related set up, the coupling of topological
matter multiplets to topological 2D gravity \cite{suit3} has been used to
investigate
non critical string theories
and relations have been established
with the integrable hierarchies discovered in matrix models \cite{suit4}.

{}From a formal field-theoretic point of view the general framework
of topological field-theories is that of geometrical
BRST-quantization \cite{suit5}. One
deals with a classical Lagrangian that has a very large symmetry,
such as the group of continuous deformations of a gauge-connection
or of a metric and which, therefore, is a topological-invariant-density
(i.e. some characteristic class of some fibre-bundle). To this symmetry
one applies the standard BRST quantization scheme and, in this way, one
obtains a topological BRST-cohomology, namely a double elliptic
complex involving both the standard exterior derivative $d^2=0$
and a second nilpotent operator (the Slavnov operator $s^2=0$) that
anticommutes with the first: $sd+ds=0$. The true geometrical and physical
content of the theory emerges when one fixes the gauge: indeed the gauge fixing
condition is, normally, some kind of self-duality condition that reduces the
space of physical states to the space of suitable {\sl instantons}.

In this perspective  the relevance of the topological twist is appreciated.
This is a procedure, discovered by Witten \cite{witten}, that extracts a
topological
field-theory with its gauge already fixed to a suitable instanton condition
from an
N=2 supersymmetric ordinary field-theory. Actually the very first example of
topological field-theory, namely
Donaldson theory, was constructed in this way
starting from N=2 super Yang-Mills theory.  The basic ingredients of the twist
procedure are:

i) the possibility of changing the spins of the fields, by redefining
a new Lorentz group as the diagonal of the old one (or a factor thereof)
with an internal symmetry group, in such a way that, after the twist,
the top spin boson of each supersymmetric multiplet and one of its
fermionic partners acquire the same spin in the new theory;

ii) the existence of an additional U(1)-symmetry of the old theory, such that,
redefining also the ghost number as the old one  plus this particular
U(1)-charge, the anticommuting partners of the bosons, that have acquired the
same spin in the twist procedure, have, in the new theory,  ghost number one,
while their bosonic partners remain with ghost number zero. In this way the
old fermions become the ghost associated with the topological symmetry.

The twist not only provides a constructive procedure for topological
field-theories but also illuminates  the topological character of a sector of
the
parent theory.  This way of thinking has been most successfully implemented
in two-dimensions. There the (Euclidean) Lorentz group is SO(2) and it can
be easily redefined by taking its diagonal with the U(1) automorphism group
of N=2 supersymmetry. In this simple case, the same U(1) provides also the
charge to shift the ghost numbers. The result, as already mentioned, is  given
by either the topological sigma-models, or the topological Landau-Ginzburg
models, or their coupling to topological 2D gravity. The topological sector
of the original N=2 theory that is unveiled by this twist procedure is that of
the chiral correlation functions.
\par
In four-dimensions the twist procedure relies once more on the properties of
N=2 supersymmetry, but involves many more subtleties, so that the programme of
topologically twisting all N=2, D=4 theories needs deeper thinking.  This
programme has been started in \cite{anselmifre} by twisting pure N=2
supergravity: in the present
paper we push this programme one step further by twisting
N=2 supergravity coupled to vector multiplets and  by discussing the effect
of the twist on N=2 hypermultiplets. The accomplished result of the present
paper is given by a  D=4 topological  Yang-Mills theory coupled to topological
D=4 gravity, the space of physical states being the moduli-space
of gauge-instantons living in the background of gravitational instantons.  One
of the properties of this theory is that it does not seem to reduce to
Donaldson
theory in the limit where the gravitational coupling is switched off. Hence
it seems to define a different topological Yang-Mills theory. Whether this
difference is substantial or not is still to be clarified; anyhow it is not
accidental rather it is deeply rooted in the properties of N=2 supersymmetry.
\par
Indeed the subtleties one encounters in twisting N=2,D=4 theories
relate mostly to the second item of the twisting programme, namely  to the
identification of the U(1) symmetry needed  to shift the ghost-number. This
identification is involved with the non-linear sigma model structure
of the original N=2 theory, in particular with the special
K\"ahler geometry of the vector multiplet coupling.  In this paper we find out
that the required U(1)-symmetry, named by us  R-duality, exists,  in the
supergravity coupled case,  if the Special K\"ahler manifold is chosen to be
$SU(1,n)/SU(n) \times U(1)$, the so named minimal coupling case. In the
flat case the  needed U(1) also exists, as Witten construction shows, if the
minimal coupling is selected. The point is that  the minimal coupling in
flat space and in curved space correspond to different unequivalent
sigma model geometries: the flat $C^n$-manifold versus the special
K\"ahler manifold $SU(1,n)/SU(n) \times U(1)$. This shows how the
flat space limit of the gravity coupled topological Yang-Mills theory is
in principle different from Donaldson theory as constructed by Witten.

Other subtleties of the D=4 topological twist were already encountered
and resolved in our previous paper on pure N=2 supergravity. Indeed
the greater complexity of N=2 supergravity with respect to N=2 super Yang-Mills
forced us \cite{anselmifre} to generalize the procedure of topological twist
as introduced by Witten \cite{witten} in N=2 super Yang-Mills and
at the same time lead us to reach a deeper understanding of  its structure.
In particular, we stress that the twist acts only on
the Lorentz indices and not on  the space-time indices \cite{anselmifre}
and this is  quite natural in the formalism  of differential forms.
This feature of the twist
avoids the problem encountered by Witten in Ref.\ \cite{witten},
namely that the twisting procedure is meaningful only when space-time is
$R^4$. We shall come back on this aspect extensively in this paper.
When one studies the topological sector of N=2 matter coupled supergravity,
one soon realizes that other aspects of the twist
still need a better understanding.
In particular,  as we already pointed out, the fundamental question is the
following: what is, in general  terms, the $U(1)$ symmetry that leads to the
ghost number of the topological version of a  given theory? In N=2 super
Yang-Mills,
as well as in N=2 pure supergravity there is only one $U(1)$ internal
symmetry (apart from global dimensional
rescalings, that are not relevant to our
discussion)
and so either it works or not. Fortunately it works. However,
in N=2 supergravity coupled to vector multiplets, there can be more that
one internal $U(1)$; think for example of
the $U(1)$ K\"ahler transformation or some $U(1)$
subgroup of the group of duality transformations \cite{gaillardzumino}
(at least when the vectors are not gauged). Anyway neither of these
two known possibilities has the correct properties to become a ghost
number and further on
we show that indeed they cannot do the job.
On the other hand one expects that a twist is possible, since the theory
of topological gravity coupled to topological Yang-Mills should
exist. In Ref.\ \cite{anselmifre} we have shown how to produce a
gauge-free algebra and generic observables
for {\sl any} topological theory and it would be very surprising to find
that it is impossible to choose any kind of instantons to fix the topological
symmetry and a gauge fermion to give a lagrangian to the theory. So, we
start our work with the belief that if a suitable $U(1)$ internal charge
is missing, this is because it is not known and not because it does not exist.
As anticipated,
it will be named R-duality, for reasons that we shall explain.
First we define it
and this lead us to single out the basic properties
an internal $U(1)$ symmetry should have
in order to give ghost number. Then
we shall explicitly
prove invariance of the minimally coupled theory
under this symmetry.

Our paper is organized as follows.
In section II we make some general
remarks on the possibility that minimal N=2 matter coupled
supergravity possesses the desired internal $U(1)$ symmetry (R-duality).
In section III we recall the structure of N=2 matter coupled supergravity
in the rheonomy framework. In section IV we fully determine R-duality
and prove that it is indeed an on shell symmetry of the theory. In section V
we present the topologically twisted-topologically shifted theory
(the gauge-free algebra, the complete BRST algebra,
the topological gauge-fixings, the observables, the gauge-fermion).
Finally, in section VI we discuss the twist of quaternionic matter multiplets
coupled to N=2 supergravity and along with this discussion, we
summarize all the steps of the twisting procedure in four dimensions,
improved by the experience of the present paper.

\section{General remarks on R-duality}
\label{general}

In this section we discuss the possibility that
minimal N=2 matter coupled supergravity is
R-duality invariant.
This internal $U(1)$ charge will add to the ghost number to define
the ghost number of the topologically twisted theory.
Thus we shall be able to extend the procedure of topological twist
and topological shift of Ref.\ \cite{anselmifre} in a rather direct way.

Let us first make some
simple remarks about the properties of the
chiral-dual invariance displayed by N=2 simple supergravity.
These properties
will guide us in finding the desired generalization to the
matter coupled case. We use the same notation
of Ref.\ \cite{anselmifre}. Consider the Bianchi identity of
the graviphoton $A$, that is
\begin{equation}
{\cal D}\!R^{\otimes}+2\epsilon_{AB}\bar\psi_A\wedge\rho_B=0,
\end{equation}
its equation of motion,
\begin{equation}
4i\epsilon_{AB}\bar\rho_A\wedge \gamma_5\psi_B-{\cal D}(F^{ab}V^c\wedge V^d)
\epsilon_{abcd}=0,
\label{eqmo}
\end{equation}
the rheonomic parametrization of the graviphoton
curvature $R^\otimes$,
\begin{equation}
R^\otimes=F_{ab}V^a\wedge V^b,
\end{equation}
and the on shell chiral-dual transformation, i.\ e.\
\begin{eqnarray}
\hat\delta \psi_A=i\gamma_5\psi_A\nonumber\\
\hat\delta F_{ab}=-2i\tilde F_{ab}=\epsilon_{abcd}F^{cd}.
\label{chidua}
\end{eqnarray}
In Ref.\ \cite{anselmifre} it was noted that the chiral-dual variation
of the Bianchi identity is the equation of motion and {\sl viceversa}. This is
evident if we re-write the Bianchi identity of the graviphoton and its equation
of
motion in the following form
\begin{eqnarray}
d[R^\otimes-\epsilon_{AB}\bar\psi_A\wedge\psi_B]&=&0,\nonumber\\
d[\epsilon_{abcd}F^{ab}V^c\wedge V^d-
2i\epsilon_{AB}\bar\psi^A\wedge\gamma_5\psi^B]&=&0.
\end{eqnarray}
Moreover, let us see what is the condition for the transformation
(\ref{chidua}) to be well defined,
i.\ e.\ what is required for the existence of
a $\hat\delta A$ compatible with (\ref{chidua}).
One immediately finds
\begin{eqnarray}
\epsilon_{abcd}F^{cd}V^a\wedge V^b&=&\hat\delta[F_{ab}V^a\wedge
V^b]=\nonumber\\
=\hat\delta R^\otimes&=&d\hat\delta
A+2i\epsilon_{AB}\bar\psi_A\wedge\gamma_5\psi_B.
\end{eqnarray}
So, $\epsilon_{abcd}F^{cd}V^a\wedge V^b-2i\epsilon_{AB}\bar\psi_A\wedge
\gamma_5\psi_B$
must be an exact form and we focus on the case
in which a necessary and sufficient condition
for this to be true
is that the form is closed, i.\ e.\
$d[\epsilon_{abcd}F^{cd}V^a\wedge V^b-2i\epsilon_{AB}\bar\psi_A\wedge\gamma_5
\psi_B]=0$. This is precisely the equation of
motion for the graviphoton (\ref{eqmo}). Consequently, the
$U(1)$ transformation is defined on shell and only on shell.
This way of reasoning is a natural generalization of the
well known case of electromagnetism and it will directly extend
to N=2 matter coupled supergravity.

What do we expect R-duality to be like? Obviously, it should
reduce to the known results both on the gravitational multiplet when matter is
suppressed and on the vector multiplets when gravity is switched off.
In other words, it should be a dual transformation on the graviphoton
(that is why we call it {\sl duality}),
a chiral transformation on the fermions and should leave the graviton and
the matter vectors inert. The scalars of the vector multiplets should
have charges $+2$ and $-2$. Consequently, on the fields of the
vector multiplets the symmetry we are
seeking should act as the usual internal $U(1)$ symmetry of
N=2 super Yang-Mills, which is an R-symmetry \cite{fayetferrara}.
Finally, it should be possible to gauge
the matter vectors (but not the graviphoton) while preserving the symmetry.

We expect R-duality not to be present in the most general case,
i.\ e.\ with any special K\"ahler manifold, but only in the simplest case,
namely for
minimal coupling \cite{dewit}. This is suggested by the fact
that something similar seems to
happen even in the case of flat N=2 super Yang-Mills theory.
As a matter of fact, the theory involves
the choice of an arbitrary flat special geometry prepotential $F(X)$,
which is a holomorphic homogeneous function of degree two
of the simplectic sections $X_\Lambda$ \cite{cremmer}.
As a result, the lagrangian involves a coupling matrix
$f^{ij}(z)$, which, in flat coordinates $z_i={X_i\over X_0}$,
depends holomorphically on the scalars $z_i$
and is given by the second derivative of $F$, $f^{ij}(z)=
{\partial\over \partial z_i}
{\partial\over \partial z_j}F(X(z))$ \cite{cremmer}.
The kinetic lagrangian of the vectors has the following form
\begin{equation}
F^i_{\mu\nu}F^{j\mu\nu}{\rm Re} \hskip .05truecm f^{ij}-
{1\over 2}\varepsilon^{\mu\nu\rho\sigma}
F^i_{\mu\nu}F^j_{\rho\sigma}{\rm Im} \hskip .05truecm f^{ij}.
\end{equation}
Only when $f^{ij}(z)=\delta^{ij}$,
namely when $F$ is quadratic, there is an evident R-invariance, since if $z$
has a nonvanishing charge, then
the only neutral holomorphic function of $z$ is the constant.
In other words, the topological twist appears to be
possible only in one case,
although the negative result that R-symmetry is barred in nonminimal
coupling has not been established in a conclusive way.
Indeed, we shall prove that R-duality exists in minimal matter
coupled N=2 supergravity, but we shall not prove
that this is the only possible case. There could
be some unexpected field redefinitions that make it work in more general
cases, even if they presumably cannot
make it suitable for a topological twist.
Uniqueness remains, for the time being, just our conjecture.

We recall that in topological Yang-Mills theory the chiral anomaly
becomes ghost number anomaly after the twist and can be described by
saying that the functional measure has a definite nonvanishing
ghost number. Consequently, only the amplitudes of observables that
have a total ghost number opposite to this value
can be nontrivial. These features of
ghost number are present also in topological gravity with or
without matter. In Ref.\ \cite{dolgov} it is shown that
the dual invariance of Maxwell theory in external gravity is anomalous.
In topological gravity
we thus expect a ghost number anomaly which is due not only to the
anomalous chiral behaviour of the fermions, but also to the
anomalous dual behaviour
of the graviphoton. In other words one has to take care of
the zero modes of the graviphoton, besides those of the fermions.

Let us now derive some {\sl a priori} information about R-duality.
As in Ref.\ \cite{anselmifre} to each field of the theory we assign
a set of labels $^c(L,R,I)^g_f$,
where $L$ is the representation of $SU(2)_L$,
$R$ is the representation of $SU(2)_R$,
$I$ is the representation of $SU(2)_I$, $c$ is the $U(1)_I$ charge,
$g$ the ghost number and $f$ the form number.
If the twist acts on $SU(2)_R$,
then after the twist we have objects described by
$(L,R\otimes I)^{g+c}_f$. In this case the left handed
components of gravitinos and gauginos must necessarily have $U(1)_I$ charge
$+1$, since they are the only fermions that have the correct spin content
to give the topological ghosts after the twist.
For example, the left handed components of the gravitinos are characterized by
$({1\over 2},0,{1\over 2})^0_1$ and give $({1\over 2},{1\over 2})_1$
after the twist,
and the vierbein $V^{a}$ is also a $({1\over 2},{1\over 2})_1$ object.
Similarly, the left handed components of the gauginos become
$({1\over 2},{1\over 2})_0$ after the twist: let us call them $\lambda_a$.
The vector bosons,
however, are Lorentz scalars, so they give $(0,0)^0_1$. Consequently, the
correct topological ghosts can only be $\lambda_a V^a$.

The charge of the right handed components of gravitinos and gauginos
is fixed to be
$-1$ by the fact that they are the natural candidates
to become the topological antighosts,
as far as  their Lorentz
transformation properties
are concerned. As a check, we can also see that
the charge of the right handed gravitinos is independently fixed by the
following argument to the value $c=-1$.
The supersymmetry charges must also transform. In fact,
the right handed components of the supersymmetry ghosts,
which are the ghost partners of the right handed gravitinos
and so must have the same charge,
are characterized by $(0,{1\over 2},{1\over 2})^1_0$ and give
$(0,1)_0\oplus (0,0)_0$ after the twist. This is the only possibility
to obtain a scalar zero form from the supersymmetry ghosts and we recall
\cite{anselmifre} that the $(0,0)_0$ component must be topologically shifted
by a constant in order to define the BRST symmetry of the topological theory.
This implies $g+c=0$ for the right handed components of the supersymmetry
ghosts, and so $c=-1$.

We conclude that on any of the so far considered
fermions, collectively denoted by $\lambda$ (supersymmetry
ghosts included),
R-duality acts as follows
\begin{eqnarray}
\hat\delta \lambda_L=\lambda_L\nonumber\\
\hat\delta \lambda_R=-\lambda_R,
\end{eqnarray}
where $\hat\delta$ denotes R-duality and $\lambda_L$, $\lambda_R$ are the left
and right handed components, respectively. This automatically rules out the
$U(1)$ K\"ahler transformation as a candidate for R-duality,
since the $U(1)$ K\"ahler charges of the gaugino
and gravitino left handed components are
opposite to each other \cite{dauriaferrarafre}. Note that the
previous reasonings are not applicable to the case of
hypermultiplets. Indeed, we shall find that the  left handed
components of the spinors contained in these multiplets
have charge $-1$, while the
right handed ones have
charge $+1$ (Section \ref{sectquater}).

Once we have fixed the charges of the fermions,
the R-duality transformations
of the bosons are uniquely
fixed by requiring on shell consistency with supersymmetry,
$\delta_\varepsilon$, i.\ e.\
\begin{equation}
[\hat\delta,\delta_\varepsilon]=0.
\end{equation}
Before giving the complete result obtained from this requirement, we recall
the structure of N=2 matter coupled supergravity.

\section{N=2 supergravity plus vector multiplets in the minimal coupling case}
\label{sugra}

By definition, N=2 supergravity minimally coupled to $n$ vector multiplets
corresponds to the case where the special K\"ahler
manifold spanned by the vector multiplet scalars
is the homogeneous manifold
${\cal M}={SU(1,n)\over SU(n)\otimes U(1)}$. In the language
of holomorphic prepotentials this corresponds to the choice
$F(X)={1\over 4}({X_0}^2-\sum_{i=1}^n {X_i}^2)$.
An easy way to obtain the explicit
form of this theory, in the rheonomy framework
that we use throughout the paper,
is by truncation of N=3
matter coupled supergravity \cite{castdauriafre,ceresole}.
If we are interested in
the case of just one vector multiplet, it is more convenient to
truncate pure N=4 $SO(4)$ supergravity \cite{cremmerscherk}. As a matter of
fact, we first tested our conjectures using this trick
(which we do not discuss here) and, after
having found that their were correct, we extended them to
$n$ vector multiplets in the way we now present.

The gravitational multiplet is $(V^a,\psi_A,\psi^A,A_0)$
(the index $A$ taking the values $1,2$),
where $V^a$ is the vierbein, $\psi_A$ are the
gravitino left handed components
($\gamma_5 \psi_A=\psi_A$),
$\psi^A$ are the right handed ones ($\gamma_5 \psi^A=-\psi^A$)
and $A_0$ is the graviphoton. The $n$ vector multiplets are labelled by an
index $i=1, \ldots n$ and are denoted by
($A_i,\lambda^A_i,\lambda^i_A,z_i,\bar z^i)$, $A_i$ being the vector bosons,
$\lambda^A_i$ the gaugino left handed components,
$\lambda_A^i$ the right handed ones, $z_i$ and $\bar z^i$ the scalars.
Vierbein, gravitinos, graviphoton and vector bosons are 1-forms,
all the other fields being 0-forms.

A special K\"ahler manifold $SK(n)$
is a Hodge K\"ahler manifold providing the base manifold for
a flat $Sp(2n+2)$ simplectic vector bundle
 ${\cal S}\stackrel{\pi}{\rightarrow} SK(n)$, whose holomorphic sections
$(X_\Lambda,{\partial F\over \partial X_\Lambda})$, $\Lambda=
0,1\ldots n,$ are given in terms of a prepotential $F(X)$,
homogeneous of degree two in the $n+1$ variables $X_\Lambda(z)$
($z$ belonging to $SK(n)$).
It is common to introduce the following expressions
\begin{eqnarray}
F^{\Lambda \Sigma}&=&\partial^\Lambda\partial^\Sigma F(X),\nonumber\\
N^{\Lambda \Sigma}&=&F^{\Lambda \Sigma}+\bar F^{\Lambda \Sigma},\nonumber\\
G&=&-{\rm ln}(N^{\Lambda\Sigma}X_\Lambda\bar X_\Sigma),\nonumber\\
L_\Lambda&=&e^{G\over 2}X_\Lambda,\nonumber\\
f_\Lambda^i&=&\partial^iL_\Lambda
+{1\over 2} G^i L_\Lambda,\nonumber\\
{\cal N}^{\Lambda \Sigma}&=&-\bar F^{\Lambda \Sigma}+{1\over
N^{\Delta\Gamma}L_\Delta L_\Gamma}N^{\Lambda\Pi}L_\Pi N^{\Sigma \Xi}L_\Xi,
\label{definitions}
\end{eqnarray}
where $G$ is the K\"ahler potential,
$\partial^\Lambda={\partial\over \partial X_\Lambda}$,
$\partial^i={\partial\over \partial z_i}$,
$G^i=\partial^i G$.

In the minimal case, if we use the special
coordinates $z_\Lambda={X_\Lambda\over X_0}$ ($z_0=1$) and furthermore
we impose $X_0\equiv 1$, then $F(z)={1\over 4}(1-\sum_{i=1}^n
z_i z_i)$ and
\begin{eqnarray}
F^{\Lambda \Sigma}&=&={1\over 2}\eta^{\Lambda\Sigma}=
{1\over 2}{\rm diag}(1,-1,\ldots -1),
\nonumber\\
N^{\Lambda \Sigma}&=&\eta^{\Lambda\Sigma},\nonumber\\
G&=&-{\rm ln}a,\nonumber\\
L_\Lambda&=&{z_\Lambda\over \sqrt{a}},\nonumber\\
f_\Lambda^i&=&\left(\matrix{f^i_0\cr
                            f^i_j}\right)={1\over a\sqrt{a}}
              \left(\matrix{\bar z^i\cr
                            a\delta^i_j+z_j\bar z^i}\right),\nonumber\\
{\cal N}^{\Lambda \Sigma}&=&\left(\matrix{{\cal N}^{00}&{\cal N}^{0j}\cr
                                          {\cal N}^{i0}&{\cal N}^{ij}}\right)=
{1\over 2(1-z_i z_i)}\left(\matrix{1+z_l z_l & -2 z_j\cr
                                -2z_i & \delta_{ij}(1-z_lz_l)+2z_iz_j}\right),
\label{minimaldefinitions}
\end{eqnarray}
where $a=1-z_i\bar z^i$.

In the notation of N=3 matter coupled supergravity
\cite{castdauriafre,ceresole},
the manifold ${{\cal G}\over{\cal H}}={SU(3,n)\over SU(3)\otimes SU(n)\otimes
U(1)}$ (which becomes ${\cal M}={SU(1,n)\over SU(n)\otimes
U(1)}$ when truncating to N=2), is described by
a matrix ${L_\Lambda}^\Sigma(z,\bar z)$ that depends on the coordinates
$z_i^A,\bar z_i^A\equiv z^i_A$,
where $A=1,2,3$, $i=1,\ldots n$, $\Lambda=(A,i)$. The N=2 truncation
is realized by setting to zero the fermions that have index $A=3$,
the bosons with $A=1,2$, the spin 1/2 of the N=3 graviton multiplet
and the $SU(3)$-singlet spin 1/2 fields of the vector multiplets.
The $L$ matrix is \cite{castdauriafre,ceresole}
\begin{equation}
{L_\Lambda}^\Sigma(z,\bar z)=\left(\matrix{
                 {L_1}^1 & {L_1}^2&{L_1}^3&{L_1}^j\cr
                 {L_2}^1 & {L_2}^2&{L_2}^3&{L_2}^j\cr
                 {L_3}^1 & {L_3}^2&{L_3}^3&{L_3}^j\cr
                 {L_i}^1 & {L_i}^2&{L_i}^3&{L_i}^j}\right)={1\over
\sqrt{a}}\left(\matrix{
                 1&0&0&0\cr
                 0&1&0&0\cr
                 0&0&1 & \bar z^j\cr
                 0&0&z_i & {M_i}^j}\right),
\end{equation}
where ${M_i}^j=\sqrt{a}\delta_i^j+{z_i\bar z^j\over |z|^2}(1-\sqrt{a})$.
The correspondence with the N=2 notation is the following
\begin{equation}
{L_\Lambda}^\Sigma=\left(\matrix{
                         1&0&0&0\cr
                         0&1&0&0\cr
                         0&0&L_0 & f_0^k {(g^{-{1\over 2}})_k}^j\cr
                         0&0&L_i & f_i^k {(g^{-{1\over 2}})_k}^j}\right),
\end{equation}
where ${(g^{-{1\over 2}})_i}^j=\sqrt{a}\delta_i^j+
{z_i\bar z^j\over |z|^2}(a-\sqrt{a})$. Note that
${{1\over a}M_i}^k {{1\over a}M_k}^j={g_i}^j
\equiv \partial_i \partial^j G$, where $\partial_i={\partial^i}^*$;
${g_i}^j$ is the metric tensor
of the K\"ahler manifold $\cal M$. We thus define ${{1\over a}M_i}^j=
{(g^{1\over 2})_i}^j$, and  $a{{M^{-1}}_i}^j=
{(g^{-{1\over 2}})_i}^j$.

The N=2 truncation of the ${\cal G}\over {\cal H}$ connection
${\Omega_\Lambda}^\Sigma$ is
\begin{equation}
{\Omega_\Lambda}^\Sigma={(L^{-1})_\Lambda}^\Pi (d{L_\Pi}^\Sigma+g {f_\Pi}^
{\Delta \Gamma} A_\Delta {L_\Gamma}^\Sigma)\equiv
\left(\matrix{0&0&0&0\cr
              0&0&0&0\cr
              0&0&-i Q & P^j\cr
              0&0&P_i & Q_i^j+{i\over n}\delta_i^j Q}\right).
\label{piequation}
\end{equation}
In particular, $Q$ is the gauged K\"ahler connection and $P^i$ is the gauged
vierbein on $\cal M$,
\begin{eqnarray}
Q&=&-{i\over 2}(G^i\nabla z_i-G_i\nabla\bar z^i),\nonumber\\
P_i&=&{(g^{1\over 2})_i}^j\nabla z_j
\label{extra}
\end{eqnarray}
and $P^i=P_i^*$.
{}From now on, let $\Lambda$ take only the values $(A=3,i=1,\ldots n)$. For
convenience, the index $3$ will be eventually replaced by a $0$
or simply omitted, when there can be no misunderstanding.

At this point, truncating the N=3 curvature definitions (see
Eq.s (IV.7.46) and (IV.7.48) of Ref.\ \cite{castdauriafre}),
we obtain the N=2 curvature definitions already adapted to the minimal
coupling.

\begin{eqnarray}
R^a&=&dV^a-\omega^{ab}\wedge V_b-i\bar\psi_A\gamma^a\wedge\psi^A\equiv
{\cal D}V^a-i\bar\psi_A\wedge\gamma^a\psi^A,\nonumber\\
R^{ab}&=&d\omega^{ab}-\omega^{ac}\wedge {\omega_c}^b,\nonumber\\
\rho_A&=&d\psi_A-{1\over 4}\omega^{ab}\gamma_{ab}\wedge\psi_A+
{i\over 2}Q\wedge\psi_A=
{\cal D}\psi_A+{i\over 2}Q\wedge\psi_A\equiv\nabla\psi_A,\nonumber\\
\rho^A&=&d\psi^A-{1\over 4}\omega^{ab}\gamma_{ab}\wedge\psi^A-
{i\over 2}Q\wedge\psi^A=
{\cal D}\psi^A-{i\over 2}Q\wedge\psi^A\equiv\nabla\psi^A,\nonumber\\
F_\Lambda&=&dA_\Lambda+{f_\Lambda}^{\Omega\Delta}A_\Omega\wedge A_\Delta+
\epsilon_{AB}L_\Lambda \bar\psi^A\wedge\psi^B+\epsilon^{AB}\bar L_\Lambda
\bar\psi_A\wedge\psi_B,\nonumber\\
\nabla\lambda_{iA}&=&d\lambda_{iA}-{1\over 4}\omega^{ab}\wedge\gamma_{ab}
\lambda_{iA}+{i\over 2}\left(1+{2\over n}\right)Q
\lambda_{iA}+{Q_i}^j\lambda_{jA},\nonumber\\
\nabla\lambda^{iA}&=&d\lambda^{iA}-{1\over 4}\omega^{ab}\wedge\gamma_{ab}
\lambda^{iA}-{i\over 2}\left(1+{2\over n}\right)Q \lambda^{iA}
+{Q^i}_j\lambda^{jA},\nonumber\\
\nabla z_i&=&dz_i+gA_\Lambda k^\Lambda_i(z),\nonumber\\
\nabla \bar z^i&=&d\bar z^i+gA_\Lambda k^{i\Lambda}(\bar z),
\label{curvatures}
\end{eqnarray}
where $\gamma_{ab}={1\over 2}[\gamma_a,\gamma_b]$ and ${Q^i}_j=({Q_i}^j)^*$.
$k_{\Lambda i}(z)$ and $k^i_{\Lambda}(\bar z)$ are respectively
the holomorphic
and antiholomorphic Killing vectors generating the special
K\"ahler manifold isometries.
The explicit expression of these Killing vectors can be read from Eq.s
(\ref{piequation}) and (\ref{extra}), isolating the
term
proportional to $A_\Lambda$ in the definition of
$P_i={(g^{1\over 2})_i}^j(dz_j+gA_\Lambda k_j^\Lambda(z))$.
One finds $k_i^\Lambda(z)={f_i}^{\Lambda k}z_k$ in the
case in which only the matter vectors are gauged (this
point will be justified in the following section).
In the N=2 notation it is useful to introduce the new definitions
\begin{eqnarray}
\lambda_i^A&=&-\epsilon^{AB}{(g^{-{1\over 2}})_i}^j\lambda_{jB},\nonumber\\
\lambda^i_A&=&-\epsilon_{AB}{(g^{-{1\over 2}})_j}^i\lambda^{jB}.
\label{redef}
\end{eqnarray}
Since $z$ and $\bar z$ will be shown to have
opposite R-duality charges, the matrix $g^{1\over 2}$ is R-duality invariant
and so
the above definitions do not change the R-duality transformation
properties of the fermions. Formulae (\ref{redef}) are determined in such
a way as to match the following rheonomic parametrizations
\begin{eqnarray}
P_i&=&P_{i|a}V^a+\epsilon^{AB}\bar\lambda_{iA}\psi_B,\nonumber\\
\nabla z_i&=&Z_{i|a}V^a+\bar\lambda_i^A\psi_A,
\label{rheo1}
\end{eqnarray}
that appear in the N=3 and N=2 formulations, respectively.
In the N=2 notation the gaugino curvatures are
\begin{eqnarray}
\nabla \lambda_i^A&=&{\cal D}\lambda_i^A-{i\over 2}Q\lambda^A_i
-{\Gamma_i}^j\lambda_j^A,\nonumber\\
\nabla \lambda^i_A&=&{\cal D}\lambda^i_A+{i\over 2}Q\lambda_A^i
-{\Gamma^i}_j\lambda^j_A,
\end{eqnarray}
where ${\Gamma_i}^j=-{(g^{-1})_i}^l(\partial^j {g_l}^k)\nabla z_k
-gA_\Lambda\partial^j k_i^\Lambda$
is the gauged Levi-Civita holomorphic connection on $\cal M$
and ${\Gamma^i}_j=({\Gamma_i}^j)^*$.

In the variables $\lambda_{iA}$, $P_i$ inherited from the N=3 truncation,
the standard N=2 Bianchi identities
(see Eq.s (3.35) of Ref.\ \cite{dauriaferrarafre})
take the following form
\begin{eqnarray}
{\cal D}R^a&+&R^{ab}\wedge V_b+i\bar \rho^A\wedge \gamma^a\psi_A-i
\bar\psi^A\wedge \gamma^a \rho_A=0,\nonumber\\
{\cal D}R^{ab}&=&0,\nonumber\\
\nabla \rho_A&+&{1\over 4}R^{ab}\wedge\gamma_{ab}\psi_A-
{i\over 2}K\wedge\psi_A=0,\nonumber\\
\nabla \rho^A&+&{1\over 4}R^{ab}\wedge\gamma_{ab}\psi^A+
{i\over 2}K\wedge\psi^A=0,\nonumber\\
\nabla F_\Lambda&-&f_\Lambda^i \nabla z_i \epsilon_{AB}\bar\psi^A\wedge\psi^B
-\bar f_{\Lambda i} \nabla \bar z^i \epsilon^{AB}\bar\psi_A\wedge\psi_B+
\nonumber\\
&+&
2L_\Lambda\epsilon_{AB}\bar\psi^A\wedge\rho^B+
2\bar L_\Lambda\epsilon^{AB}\bar\psi_A\wedge\rho_B=0,\nonumber\\
\nabla^2 \lambda_{iA}&+&{1\over 4}R^{ab}\wedge\gamma_{ab}\lambda_{iA}-
{R_i}^j\lambda_{jA}-{i\over 2}\left(1+{2\over n}\right)K\lambda_{iA}=0,
\nonumber\\
\nabla^2 \lambda^{iA}&+&{1\over 4}R^{ab}\wedge\gamma_{ab}\lambda^{iA}-
{R^i}_j\lambda^{jA}+{i\over 2}\left(1+{2\over n}\right)K\lambda^{iA}=0,
\nonumber\\
\nabla P_i&=&dP_i+{Q_i}^j\wedge P_j+i\left(1+{1\over n}\right)
Q\wedge P_i=0,\nonumber\\
\nabla P_i&=&dP^i+{Q^i}_j\wedge P^j-i\left(1+{1\over n}\right)
Q\wedge P^i=0,
\label{bianchi}
\end{eqnarray}
where $K=dQ$, ${R_i}^j=d{Q_i}^j+{Q_i}^k\wedge{Q_k}^j$ and
${R^i}_j=({R_i}^j)^*$.
The rheonomic parametrizations are
\begin{eqnarray}
R^a&=&0,\nonumber\\
R^{ab}&=&{R^{ab}}_{cd}V^c\wedge V^d-i\bar\psi_A(2\gamma^{[a}\rho^{A|b]c}
-\gamma^c\rho^{A|ab})\wedge V_c+\nonumber\\
&-&i\bar\psi^A(2\gamma^{[a}{\rho_A}^{|b]c}
-\gamma^c{\rho_A}^{|ab})\wedge V_c+2G^{-ab}\epsilon^{AB}\bar\psi_A\wedge\psi_B+
\nonumber\\
&+&2G^{+ab}\epsilon_{AB}\bar\psi^A\wedge\psi^B
+{i\over 4}\varepsilon^{abcd}\bar\psi_A\wedge\gamma_c\psi^B
(2\bar\lambda_{iB}\gamma_d\lambda^{iA}-
\delta^A_B\bar\lambda_{iC}\gamma_d\lambda^{iC}),\nonumber\\
\rho_A&=&\rho_{A|ab}V^a\wedge V^b-2i \epsilon_{AB}G^+_{ab}\gamma^a\psi^B
\wedge V^b+{i\over 4}\psi_B\bar
\lambda^{iB}\gamma^a\lambda_{iA}\wedge V_a+\nonumber\\
&+&{i\over 8}\gamma_{ab}\psi_B\left(2\bar
\lambda^{iB}\gamma^a\lambda_{iA}-
\delta^B_A\bar
\lambda^{iC}\gamma^a\lambda_{iC}\right)\wedge V^b,\nonumber\\
\rho^A&=&\rho^A_{ab}V^a\wedge V^b-2i \epsilon^{AB}G^-_{ab}\gamma^a\psi_B
\wedge V^b+{i\over 4}\psi^B\bar
\lambda_{iB}\gamma^a\lambda^{iA}\wedge V_a+\nonumber\\
&+&{i\over 8}\gamma_{ab}\psi^B\left(2\bar
\lambda_{iB}\gamma^a\lambda^{iA}-
\delta_B^A\bar
\lambda_{iC}\gamma^a\lambda^{iC}\right)\wedge V^b,\nonumber\\
F_\Lambda&=&F_\Lambda^{ab}V_a\wedge V_b+i(f_\Lambda^i\bar\lambda^A_i
\gamma^a\psi^B\epsilon_{AB}+\bar f_{\Lambda i}\bar\lambda_A^i
\gamma^a\psi_B\epsilon^{AB})\wedge V_a,\nonumber\\
\nabla \lambda_{iA}&=&\nabla_a\lambda_{iA}V^a+iP_{i|a}\gamma^a\psi^B
\epsilon_{AB}+G^{+ab}_i\gamma_{ab}\psi_A
+gC_i\psi_A,\nonumber\\
\nabla \lambda^{iA}&=&\nabla_a\lambda^{iA} V^a+iP^i_{|a}\gamma^a\psi_B
\epsilon^{AB}+G_{ab}^{-i}\gamma^{ab}\psi^A
+gC^i\psi^A,\nonumber\\
\nabla z_i&=&Z_{i|a}V^a+\bar\lambda_i^A\psi_A,\nonumber\\
\nabla \bar z^i&=&\bar Z^i_{|a} V^a+ \bar\lambda^i_A\psi^A.
\label{rheo2}
\end{eqnarray}
where $C^i={(L^{-1})_3}^k{L_j}^i{L_l}^3{f_k}^{jl}$ are obtained from the
N=2 truncation as particular instances of the N=3 boosted structure constants,
$C_i=({C^i})^*$, $\bar f_{\Lambda i}=
(f_\Lambda^i)^*$. $G^{+ab}$, $G^{+ab}_i$, $G_{ab}^-$ and $G_{ab}^{-i}$
are determined by the equation
\begin{equation}
{1\over 2}F^{ab}_\Lambda+{L_\Lambda}^3 G^{-ab}+{L_\Lambda}^i G_i^{+ab}+
{(L^{-1})_3}^\Pi J_{\Pi\Lambda}G^{+ab}-
{(L^{-1})_i}^\Pi J_{\Pi\Lambda}G^{i-ab}=0,
\end{equation}
where $J_{\Lambda\Pi}$ is the $SU(1,n)$-invariant metric
\begin{equation}
J_{\Lambda\Pi}=\left(\matrix{1&0\cr 0&-\delta_{ij}}\right).
\end{equation}
One finds
\begin{eqnarray}
G^{+ab}&=&-{1\over 4}\sqrt{a}(1+2\bar {\cal N})^{0\Lambda}
F^{+ab}_\Lambda,\nonumber\\
G^{+ab}_i&=&-{1\over 4}\sqrt{a}{(g^{1\over 2})_i}^j
{(1+2\bar{\cal N})_j}^\Lambda F^{+ab}_\Lambda
+{1\over a}\bar z^2 z_i G^{+ab}+{1\over 2\sqrt{a}}z_i\bar z^k F^{+ab}_k.
\label{eigenstates}
\end{eqnarray}
and $G_{ab}^-=(G_{ab}^+)^*$ and $G_{ab}^{-i}=({{G_i}_{ab}}^+)^*$.
The rheonomic parametrizations are on-shell consistent
with the Bianchi identities (\ref{bianchi}).

We can now write down the lagrangian of N=2 supergravity minimally coupled to
$n$ vector multiplets.
\begin{equation}
{\cal L}={\cal L}_{kin}+{\cal L}_{Pauli}+{\cal L}_{torsion}+{\cal L}_{4Fermi}
+\Delta {\cal L}_{gauging}+\Delta {\cal L}_{potential},
\label{lagra}
\end{equation}
where
\begin{eqnarray}
{\cal L}_{kin}&=&\varepsilon_{abcd}R^{ab}\wedge V^c\wedge V^d
-4(\bar\psi^A\wedge\gamma_a\rho_A+\bar\rho^A\wedge\gamma_a\psi_A)\wedge V^a+
\nonumber\\
&-&{i\over 3}{g_i}^j(\bar\lambda^A_j\gamma_a\nabla\lambda_A^i+
\bar\lambda^i_A\gamma_a\nabla\lambda_j^A)\wedge V_b \wedge V_c\wedge V_d
\varepsilon^{abcd}+\nonumber\\
&+&{2\over 3}{g_i}^j
[\bar Z^i_{|a}(\nabla z_j-\bar\lambda_j^A\psi_A)
+Z_{j|a}(\nabla\bar z^i-\bar\lambda^i_A\psi^A)]\wedge V_b\wedge V_c
\wedge V_d \varepsilon^{abcd}+\nonumber\\
&+&{1\over 6}(\bar {\cal N}^{\Lambda\Sigma}F^{+ab}_\Lambda
F^{+}_{\Sigma ab}+{\cal N}^{\Lambda\Sigma}F^{-ab}_\Lambda
F^{-}_{\Sigma ab}-{g_i}^j \bar Z^i_{|a}{Z_{j|}}^a)\varepsilon_{cdef}
V^c\wedge V^d\wedge V^e\wedge V^f+\nonumber\\
&-&4i(\bar {\cal N}^{\Lambda\Sigma}F^{+ab}_\Lambda-
{\cal N}^{\Lambda\Sigma}F^{-ab}_\Lambda)\wedge (F_\Sigma+\nonumber\\
&-&i(f_\Sigma^i\bar\lambda^A_i
\gamma^c\psi^B\epsilon_{AB}+\bar f_{\Sigma i}\bar\lambda_A^i
\gamma^c\psi_B\epsilon^{AB})\wedge V_c)\wedge V_a\wedge V_b,\nonumber\\
{\cal L}_{Pauli}&=&-4iF_\Lambda\wedge({\cal N}^{\Lambda\Sigma} L_\Sigma
\epsilon_{AB}\bar\psi^A\wedge\psi^B-\bar{\cal N}^{\Lambda\Sigma}\bar L_\Sigma
\epsilon^{AB}\bar\psi_A\wedge\psi_B)+\nonumber\\
&+&4F_\Lambda\wedge(\bar{\cal N}^{\Lambda\Sigma}f_\Sigma^i\bar \lambda_i^A
\gamma_a\psi^B\epsilon_{AB}-{\cal N}^{\Lambda\Sigma}
\bar f_{\Sigma i}\bar \lambda^i_A
\gamma_a\psi_B\epsilon^{AB})\wedge V^a+\nonumber\\
&-&2i{g_i}^j(\nabla z_j\wedge \bar \lambda^i_A\gamma_{ab}\psi^A-
\nabla \bar z^i\wedge \bar \lambda_j^A\gamma_{ab}\psi_A)
\wedge V^a\wedge V^b,\nonumber\\
{\cal L}_{torsion}&=&R^a\wedge V_a\wedge {g_i}^j
\bar\lambda_A^i\gamma_b\lambda_j^A
\wedge V^b,\nonumber\\
{\cal L}_{4Fermi}&=&i(W\epsilon_{AB}\bar\psi^A\wedge\psi^B\wedge
\epsilon_{CD}\bar\psi^C\wedge\psi^D-\bar W
\epsilon^{AB}\bar\psi_A\wedge\psi_B\wedge
\epsilon^{CD}\bar\psi_C\wedge\psi_D)+\nonumber\\
&-&2i{g_i}^j\bar\lambda^i_A\gamma_a
\lambda_j^B\bar\psi_B\wedge\gamma_b\psi^A\wedge V^a\wedge V^b+\nonumber\\
&+i&(W_{ij}\epsilon^{AB}\bar\lambda^i_A\gamma_a\psi_B\wedge V^a\wedge
\epsilon^{CD}\bar\lambda^j_C\gamma_b\psi_D\wedge V^b+\nonumber\\
&-&W^{ij}\epsilon_{AB}\bar\lambda_i^A\gamma_a\psi^B\wedge V^a\wedge
\epsilon_{CD}\bar\lambda_j^C\gamma_b\psi^D\wedge V^b)+\nonumber\\
&+&{1\over 18}\epsilon_{abcd}V^a\wedge V^b\wedge V^c\wedge V^d
{g_i}^j\bar\lambda^i_A\gamma^m\lambda_j^A
{g_k}^l\bar\lambda^k_B\gamma_m\lambda_l^B,
\nonumber\\
\Delta {\cal L}_{gauging}&=&{2i\over 3}g(
\bar \lambda_i^A\gamma^a\psi^B W^i_{AB}
+\bar \lambda^i_A\gamma^a\psi_B W_i^{AB})
\wedge V^b\wedge V^c\wedge V^d\varepsilon_{abcd}+\nonumber\\
&+&{1\over 6}g(M^{ij}\bar\lambda_i^A\lambda_j^B\epsilon_{AB}+
M_{ij}\bar\lambda^i_A\lambda^j_B\epsilon^{AB})\varepsilon_{abcd}
V^a\wedge V^b\wedge V^c\wedge V^d,\nonumber\\
\Delta {\cal L}_{potential}&=&-{1\over 12}g^2 {g_i}^j W^i_{AB} W_j^{AB}
\varepsilon_{abcd} V^a\wedge V^b \wedge V^c \wedge V^d,
\label{lagra2}
\end{eqnarray}
where $W=2L_\Lambda L_\Sigma {\cal N}^{\Lambda \Sigma}$,
$W^{ij}=2\bar {\cal N}^{\Lambda\Sigma}f_\Lambda^i f_\Sigma^j$ and
$W_{ij}=(W^{ij})^*$, while
$M^{ij}=k^{l\Lambda}f_\Lambda^{[i}{g_l}^{j]}$
and $M_{ij}=(M^{ij})^*$,
$W^i_{AB}=\epsilon_{AB}k^{i\Lambda}L_\Lambda$,
$W_i^{AB}=(W^i_{AB})^*$.
The lagrangian in Eq.s (\ref{lagra}) and (\ref{lagra2})
agrees with the lagrangian
(4.13) of Ref.\ \cite{dauriaferrarafre} upon suppression of
the hypermultiplets and up to ${\cal L}_{4Fermi}$ and
the second term of
$\Delta{\cal L}_{gauging}$, that were not calculated in
\cite{dauriaferrarafre}.
Indeed, the very reason why we have performed the above described N=2
truncation of
the N=3 theory was that of obtaining these terms without calculating
them explicitly. Our purpose is that of checking R-duality in the
minimal coupling, however, as a byproduct, we have also obtained the complete
form of the lagrangian of N=2 supergravity coupled to vector multiplets
for an arbitrary choice of the special K\"ahler manifold.
All the objects entering (\ref{lagra2}) have already been
interpreted in a general N=2 setup
(in which the graviphoton can be gauged).
As a matter of fact,
the N=3 theory does not admit the most general gauging
of the vectors \cite{castdauriafre,ceresole}, but it surely admits
any gauging of the matter vectors. Even if the minimal N=2 theory exists
in any case, the truncation from N=3 can only give the minimal N=2
theory in which the graviphoton is not gauged.

As promised, in the following section we define R-duality and prove
that it is indeed an on-shell symmetry of the above theory.

\section{R-duality for N=2 matter coupled supergravity}
\label{Rduality}

Now, starting form the R-duality transformation properties of the fermions,
as derived in Section \ref{general}, we determine the transformations
of the bosons by simply requiring $[\hat \delta,\delta_\varepsilon]=0$
on-shell, if $\delta_\varepsilon$ is the supersymmetry transformation with
parameters $\varepsilon$ (let $\varepsilon_A$
and $\varepsilon^A$ be the left and right handed components,
respectively). The supersymmetry transformations can be read in the
usual way
from the rheonomic parametrizations (\ref{rheo1}) and (\ref{rheo2}). In any
case, their explicit expression will be written down later on in the context
of the BRST-quantization of the theory (see formula (\ref{brstalgebra})).
So, we start from
\begin{equation}
\matrix{\hat\delta \psi_A=\psi_A, & \hat\delta \varepsilon_A=\varepsilon_A, &
\hat\delta \lambda^A_i=\lambda_i^A, \cr
\hat\delta \psi^A=-\psi^A, & \hat\delta \varepsilon^A=-\varepsilon^A, &
\hat\delta \lambda_A^i=-\lambda^i_A.}
\end{equation}
First of all, consistency of R-duality
with supersymmetry states that, if a field $\phi$ has an R-duality
charge equal to $q$, then $\delta_\varepsilon \phi$ has the same charge $q$
and viceversa. It is immediate to see that $\hat \delta
\delta_\varepsilon V^a=0$ and so we deduce $\hat \delta V^a=0$.
This is good, because in our mind, R-duality is to become ghost number
and the vierbein should remain of zero ghost number together with
all the matter vectors.
Similarly, $\hat \delta \delta_\varepsilon z_i= 2 \delta_\varepsilon z_i$,
requiring $\hat\delta z_i=2 z_i$. An analogous reasoning gives,
when applied to $\bar z^i$, $\hat \delta \bar z^i=-2 \bar z^i$, thus
confirming that $z_i$ and $\bar z^i$ have opposite charges. This immediately
rules out the possibility that the $U(1)$ symmetry we are looking for might be
a subgroup of the group of duality transformations
\cite{gaillardzumino,castdauriafre}.
Indeed, in that case $z_i$ and $\bar z^i$
would have the same charge. This is welcome, because, if $U(1)_I$
were a subgroup of the duality group, we could not maintain the symmetry
in the presence of gauging, as, on the contrary, we expect to be able to do.
We immediately
see that the K\"ahler potential $G$ is invariant, as well as
the metric ${g_i}^j$
(note that this fact would not hold true in the nonminimal case).
It remains to find the transformation properties of the vector bosons.
Let us concentrate on the ungauged case ($g=0$) for the moment.
One can verify that $[\hat\delta,\delta_\varepsilon]\psi_A=0$ and
$[\hat\delta,\delta_\varepsilon]\psi^A=0$ imply
\begin{equation}
\hat\delta G^{\pm ab}=\pm 2 G^{\pm ab},
\label{dual1}
\end{equation}
while
$[\hat\delta,\delta_\varepsilon]\lambda_i^A=0$ and
$[\hat\delta,\delta_\varepsilon]\lambda^i_A=0$ imply
\begin{equation}
\matrix{\hat\delta G^{+ab}_i=0, & \hat \delta G^{i-ab}=0}.
\label{dual2}
\end{equation}
respectively.
Equations (\ref{dual1}) and (\ref{dual2}) form a linear system of equations in
$\hat\delta F^{\pm ab}_\Lambda$, in which the number of unknowns equals the
number of equations. The unique solution is
\begin{equation}
\matrix{
\hat\delta F^{0+ab}=4\bar{\cal N}^{0\Lambda}F^{+ab}_\Lambda,
                   & \hat \delta F^{+ab}_i=0,\cr
\hat\delta F^{0-ab}=-4{\cal N}^{0\Lambda}F^{-ab}_\Lambda,
                   & \hat \delta F^{i-ab}=0.}
\label{utility1}
\end{equation}
The graviphoton is thus transformed in a way that resembles the duality
transformations and this forbids its gauging it if we want R-duality.
Consequently, when considering the gauged case,
we must assume that only the matter vectors are
gauged, i.\ e.\ $f_\Lambda^{\Sigma\Omega}=0$ whenever one of the indices
$\Lambda,$ $\Sigma$, $\Omega$ takes the value zero. There is
no restriction, on the contrary,
on the gauge group of the matter vectors.

Let us
rewrite the rheonomic parametrization of the vectors
and the definition of their curvatures
\begin{eqnarray}
F_\Lambda&=&F_\Lambda^{ab}V_a\wedge V_b+i(f_\Lambda^i\bar\lambda^A_i
\gamma^a\psi^B\epsilon_{AB}+\bar f_{\Lambda i}\bar\lambda_A^i
\gamma^a\psi_B\epsilon^{AB})\wedge V_a,\nonumber\\
F_\Lambda&=&dA_\Lambda+{f_\Lambda}^{\Omega\Delta}A_\Omega\wedge A_\Delta+
\epsilon_{AB}L_\Lambda \bar\psi^A\wedge\psi^B+\epsilon^{AB}\bar L_\Lambda
\bar\psi_A\wedge\psi_B.
\label{utility2}
\end{eqnarray}
These expressions show that, under the above conditions on the structure
constants
$f_\Lambda^{\Sigma\Omega}$, the transformations $\hat \delta F^{+ab}_i=0$
and $\hat \delta F^{i-ab}=0$ imply $\hat \delta A_i=0$,
i.\ e.\ all the matter vectors are inert under R-duality (they will
have ghost number zero after the twist and this is good in order
to recover topological Yang-Mills theory).

Summarizing, R-duality acts on-shell as follows
\begin{eqnarray}
\matrix{\hat\delta V^a=0,&\cr
\hat\delta \psi_A=\psi_A,&
\hat\delta\psi^A=-\psi^A,\cr
\hat\delta F^{0+ab}=4\bar{\cal N}^{0\Lambda}F^{+ab}_\Lambda,&
\hat\delta F^{0-ab}=-4{\cal N}^{0\Lambda}F^{-ab}_\Lambda,\cr}\nonumber
\end{eqnarray}
\begin{eqnarray}
\matrix{
\hat\delta A_i=0&\cr
\hat\delta \lambda^A_i=\lambda^A_i,&
\hat\delta \lambda_A^i=-\lambda_A^i,\cr
\hat\delta z_i=2 z_i,&
\hat\delta \bar z^i=-2 \bar z^i.\cr}
\label{utility3}
\end{eqnarray}
One easily checks that formulas
(\ref{dual1}), (\ref{dual2}), (\ref{utility1}) and (\ref{utility3})
are still valid when all
the vectors but the graviphoton are gauged.

What about $\hat \delta A_0$? As in all duality-type transformations,
$\hat \delta A_0$ should be meaningful only on-shell (see Section
\ref{general}). In fact, (\ref{utility1}) and (\ref{utility2})
imply (using the explicit expressions (\ref{minimaldefinitions}))
\begin{eqnarray}
\hat \delta F_0^{ab} V_a\wedge V_b&=&4({{\bar{\cal N}}_0}^{\Lambda}
F^{+ab}_\Lambda-{{\cal N}_0}^{\Lambda}
F^{-ab}_\Lambda)V_a\wedge V_b=\nonumber\\
&=&\hat \delta [dA_0+\epsilon_{AB}L_0\bar\psi^A\wedge\psi^B+
\epsilon^{AB}\bar L_0\bar\psi_A\wedge\psi_B+\nonumber\\
&-&i(f^i_0\bar\lambda_i^A\gamma^a\psi^B
\epsilon_{AB}+\bar f_{0i}\bar\lambda^i_A\gamma^a\psi_B\epsilon^{AB})
\wedge V_a]=
\nonumber\\
&=&d\hat\delta A_0-4{{\cal N}_0}^\Lambda L_\Lambda
\epsilon_{AB}\bar\psi^A\wedge\psi^B +4 {{\bar{\cal N}}_0}^\Lambda\bar L_\Lambda
\epsilon^{AB}\bar\psi_A\wedge\psi_B+\nonumber\\
&-&4i({{\bar{\cal N}}_0}^\Lambda
f^i_\Lambda\bar\lambda_i^A\gamma^a\psi^B\epsilon_{AB}-
{{\cal N}_0}^\Lambda \bar f_{\Lambda i}\bar\lambda^i_A
\gamma^a\psi_B\epsilon^{AB})\wedge V_a.
\end{eqnarray}
Imposing $d^2 \hat \delta A_0=0$, we get
\begin{eqnarray}
d[({\bar{\cal N}_0}^{\Lambda}
F^{+ab}_\Lambda&-&{{\cal N}_0}^{\Lambda}
F^{-ab}_\Lambda)V_a\wedge V_b+{{\cal N}_0}^\Lambda L_\Lambda
\epsilon_{AB}\bar\psi^A\wedge\psi^B -{{\bar{\cal N}}_0}^\Lambda\bar L_\Lambda
\epsilon^{AB}\bar\psi_A\wedge\psi_B+\nonumber\\
&+&i({{\bar{\cal N}}_0}^\Lambda
f^i_\Lambda\bar\lambda_i^A\gamma^a\psi^B\epsilon_{AB}-
{{\cal N}_0}^\Lambda \bar f_{\Lambda i}\bar\lambda^i_A
\gamma^a\psi_B\epsilon^{AB})\wedge V_a]=0.
\end{eqnarray}
One can easily verify that this is the equation of motion of the graviphoton
as derived from the lagrangian (\ref{lagra}).
Furthermore, the R-duality variation
of the $A_0$ equation of motion is proportional to the $A_0$-Bianchi identity
and viceversa. It is easily checked that the other curvatures of
(\ref{curvatures}) and the remaining Bianchi identities of (\ref{bianchi})
transform correctly, so the last step in order to establish R-duality of
the theory is the proof of invariance for the remaining field equations.

The equations of motion of the vector bosons can be written in the
following form
\begin{equation}
dS^\Lambda+2f_\Delta^{\Lambda \Sigma}A_\Sigma\wedge
S^\Delta+R^\Lambda=0,
\end{equation}
where $S^\Lambda$ is, by definition, the coefficient in the lagrangian
of the field strength $F_\Lambda$, namely
\begin{eqnarray}
S^\Lambda&=&(\bar{\cal N}^{\Lambda\Sigma}
F^{+ab}_\Sigma-{\cal N}^{\Lambda\Sigma}
F^{-ab}_\Sigma)V_a\wedge V_b+{\cal N}^{\Lambda\Sigma} L_\Sigma
\epsilon_{AB}\bar\psi^A\wedge\psi^B -\bar{\cal N}^{\Lambda\Sigma}
\bar L_\Sigma
\epsilon^{AB}\bar\psi_A\wedge\psi_B+\nonumber\\
&+&i(\bar{\cal N}^{\Lambda\Sigma}
f^i_\Sigma\bar\lambda_i^A\gamma^a\psi^B\epsilon_{AB}-
{\cal N}^{\Lambda\Sigma} \bar f_{\Sigma i}\bar\lambda^i_A
\gamma^a\psi_B\epsilon^{AB})\wedge V_a,
\end{eqnarray}
and $R^\Lambda$ is the remainder that comes from
the ${\delta\over \delta A_\Lambda}$-variation of those terms that are
manifestly R-duality invariant and do not depend on the graviphoton $A_0$.
Since one can easily verify that
$\hat \delta S^\Lambda$ vanishes whenever $\Lambda\neq 0$
(to this purpose,
note that $\hat \delta(\bar{\cal N}^{i\Sigma}F_\Sigma^{+ab})=
\hat \delta({\cal N}^{i\Sigma}F_\Sigma^{-ab})=0$
and use the explicit expressions
(\ref{minimaldefinitions})), then the field equations
of the matter vectors are all R-duality invariant.

In order to prove R-duality
invariance of the remaining field equations, we note that it is not necessary
to study the entire lagrangian $\cal L$ (\ref{lagra}), because various terms
can give only contributions with the correct $\hat \delta$-transformation
properties. These are precisely the R-duality invariant terms of $\cal L$
that do not depend on $A_0$.
On the other hand, since
$\hat \delta F_0^{ab}$ depends on all the fields, we cannot neglect
a term $\Delta \cal L$
only because it is $\hat\delta$-invariant ($\hat\delta \Delta {\cal L}=0$)
if it contains $A_0$. Indeed,
if $\phi$ is a field of charge $q$ ($\hat\delta \phi=q\phi$;
we can take $\phi\neq A_0$ since the $A_0$-equation has already been
studied), then the
contributions to its field equation (i.\ e.\ ${\partial \over
\partial \phi}\Delta {\cal L}$) must have charge $-q$ in order to transform
correctly
($\hat\delta{\partial \over
\partial \phi}\Delta {\cal L}=-q{\partial \over
\partial \phi}\Delta {\cal L}$) and it must happen that
\begin{equation}
\left[\hat\delta,{\partial \over
\partial \phi}\right]\Delta {\cal L}=-q {\partial \over
\partial \phi}\Delta {\cal L}.
\end{equation}
For this to be true it is sufficient (and necessary, if
$\Delta {\cal L}$ has not a special form) to have
\begin{equation}
\left[\hat\delta,{\partial \over
\partial \phi}\right]\phi^\prime=-q {\partial \over
\partial \phi}\phi^\prime,
\end{equation}
for all fields $\phi^\prime$.
However, this is not true for $\phi^\prime=F^{ab}_0$ and so, if
$\Delta {\cal L}$ depends on $A_0$ one should analyze it explicitly.
Summarizing, it is sufficient to test R-duality invariance of the
contributions to the
field equations that come from the terms of the lagrangian
either containing $A_0$ or not $\hat\delta$-invariant. This
part of the lagrangian is given by
\begin{eqnarray}
\Delta {\cal L}&\equiv&
{1\over 6}(\bar {\cal N}^{\Lambda\Sigma}F^{+ab}_\Lambda
F^{+}_{\Sigma ab}+{\cal N}^{\Lambda\Sigma}F^{-ab}_\Lambda
F^{-}_{\Sigma ab})\varepsilon_{cdef}V^c\wedge V^d\wedge V^e \wedge V^f+
\nonumber\\
&-&4i(\bar {\cal N}^{\Lambda\Sigma}F^{+ab}_\Lambda-
{\cal N}^{\Lambda\Sigma}F^{-ab}_\Lambda)\wedge V_a\wedge V_b\wedge
(F_\Sigma+\nonumber\\
&-&i(f_\Sigma^i\bar\lambda^A_i
\gamma^c\psi^B\epsilon_{AB}+\bar f_{\Sigma i}\bar\lambda_A^i
\gamma^c\psi_B\epsilon^{AB})\wedge V_c),\nonumber\\
&-&2i{1\over \sqrt{a}}F^0\wedge(
\epsilon_{AB}\bar\psi^A\wedge\psi^B-
\epsilon^{AB}\bar\psi_A\wedge\psi_B)+\nonumber\\
&-&{2\over a\sqrt{a}}F^0\wedge(\bar z^i\bar \lambda_i^A
\gamma_a\psi^B\epsilon_{AB}-z_i\bar \lambda^i_A
\gamma_a\psi_B\epsilon^{AB})\wedge V^a+\nonumber\\
&+&{i\over a}(\epsilon_{AB}\bar\psi^A\wedge\psi^B\wedge
\epsilon_{CD}\bar\psi^C\wedge\psi^D-
\epsilon^{AB}\bar\psi_A\wedge\psi_B\wedge
\epsilon^{CD}\bar\psi_C\wedge\psi_D)+\nonumber\\
&-&{i\over a^3}(z_i z_j\epsilon^{AB}\bar\lambda^i_A\gamma_a
\psi_B\wedge V^a\wedge
\epsilon^{CD}\bar\lambda^j_C\gamma_b\psi_D\wedge V^b+\nonumber\\
&-&\bar z^i \bar z^j\epsilon_{AB}\bar\lambda_i^A\gamma_a\psi^B\wedge V^a\wedge
\epsilon_{CD}\bar\lambda_j^C\gamma_b\psi^D\wedge V^b),
\end{eqnarray}
where $W$ and $W^{ij}$ have been replaced
by their explicit expressions in terms of
$z_i$, and $\bar z^i$ and, after replacement, the manifestly
$\hat\delta$-invariant terms not containing $A_0$ have been deleted.
At this point, the check that the contributions to the field equations of the
fermions, the vierbein and the scalars transform correctly
is rather direct and we leave it to the reader. We thus conclude that

{\bf Proposition.}
{\it N=2 supergravity minimally coupled to $n$ vector multiplets
gauging an arbitrary $n$ dimensional group
(in which the graviphoton is not gauged),
is on-shell R-duality invariant\footnotemark
\addtocounter{footnote}{0}\footnotetext{\rm
Note that for an N=2 theory
without hypermultiplets, the statement that a certain
vector is not gauged is equivalent to the statement that it corresponds
to a $U(1)$ subgroup of the full gauge group.}.}

The possibility that R-duality exists also in the N=3 theory or in
more extended supergravity theories as well as the possibility to have
it in N=2 matter coupled supergravity in nonminimal cases (even
if, we presume, it might not be suitable for a topological twist)
remain open problems. Here we have restricted our attention to
that internal $U(1)$ symmetry
that was relevant to our purposes, that is the topological twist.

We have so far neglected the coupling of matter hypermultiplets to
N=2 supergravity, since it is immediately verified that the generalization
of R-duality due to the presence of
them is trivial. The scalars have $0$ charge, however
the left handed components of
fermions must have $-1$ charge and the right handed components
must have $+1$ charge, differently from the case of the other fermions
so far encountered.
The twist is by no means
trivial. As a matter of fact, it turns out that it is interesting as we
shall see at the end of this paper.

\section{Topological twist of the minimal theory}
\label{sectiontwist}

In this section, we discuss the twisted topological theory. First
of all, let us note that the gauge-free algebra (i.\ e.\ the minimal BRST
algebra, with neither antighosts nor gauge-fixings, nor Lagrange multipliers)
is simply the tensor product of the gauge-free algebras for topological gravity
and topological Yang-Mills \cite{anselmifre}, that is to say
\begin{eqnarray}
sA&=&-\nabla c-\psi,\nonumber\\
sc&=&\phi-{{1}\over{2}}\left[c,c\right],\nonumber\\
s\psi&=&\nabla\phi-\left[c,\psi\right],\nonumber\\
s\phi&=&-\left[c,\phi\right],\nonumber\\
sV^a&=&\psi^a-{\cal D}_0\varepsilon^a+\varepsilon^{ab}\wedge V_b,\nonumber\\
s\omega_0^{ab}&=&\chi^{ab}-{\cal D}_0\varepsilon^{ab},\nonumber\\
s\varepsilon^a&=&\phi^a+\varepsilon^{ab}\wedge\varepsilon_b,\nonumber\\
s\varepsilon^{ab}&=&\eta^{ab}+{\varepsilon^a}_c\wedge\varepsilon^{cb},
\nonumber\\
s\psi^a&=&-{\cal D}_0\phi^a+\varepsilon^{ab}\wedge\psi_b-\chi^{ab}\wedge
\varepsilon_b-\eta^{ab}\wedge V_b,\nonumber\\
s\phi^a&=&\varepsilon^{ab}\wedge\phi_b-\eta^{ab}\wedge\varepsilon_b,\nonumber\\
s\chi^{ab}&=&-{\cal D}_0\eta^{ab}+\varepsilon^{ac}\wedge{\chi_c}^b-\chi^{ac}
\wedge{\varepsilon_c}^b,\nonumber\\
s\eta^{ab}&=&\varepsilon^{ac}\wedge{\eta_c}^b-\eta^{ac}\wedge{\varepsilon_c}^b.
\label{gaugefree}
\end{eqnarray}
We have grouped the $n$ matter vectors $A_i$ into the column $A=(A_i)$.
Similarly, $\psi=(\psi_i)$, $\phi=(\phi_i)$ and $c=(c_i)$.
For the definitions of the other symbols, refer to Ref.\ \cite{anselmifre}.

The observables and the corresponding descent
equations can be derived from
the hatted extensions of the identities
$d\hskip .1truecm{\rm tr}[F\wedge F]=0$, $d\hskip .1truecm{\rm tr}[R\wedge
R]=0$ and $d\hskip .1truecm{\rm tr}[R\wedge \tilde R]=0$, in the usual way
\cite{anselmifre}.

The BRST algebra of N=2 matter coupled supergravity can be found as explained
in Ref.\ \cite{anselmifre}, that is to say by extending all
differential forms to ghost forms.
We report only the final result, that, together with the translation
ghosts $\varepsilon^a$, the Lorentz ghosts $\varepsilon^{ab}$,
the supersymmetry ghosts $c_A,c^A$, involves also the gauge ghosts $c^\Lambda$.
\begin{eqnarray}
sV^a&=&-{\cal D}\varepsilon^a+\varepsilon^{ab}\wedge V_b
+i(\bar\psi_A\wedge\gamma^a c^A
+\bar c_A \wedge \gamma \psi^A),\nonumber\\
s\varepsilon^a&=&\varepsilon^{ab}\wedge \varepsilon_b
+i\bar c_A\wedge\gamma^a c^A,\nonumber\\
s\omega^{ab}&=&-{\cal D}\varepsilon^{ab}
+2{R^{ab}}_{cd}V^c\varepsilon^d+i(\varepsilon_c\bar\psi_A+V_c \bar c_A)
(2\gamma^{[a}\rho^{A|b]c}
-\gamma^c\rho^{A|ab})+\nonumber\\
&+&i(\varepsilon_c \bar\psi^A+ V_c \bar c^A)
(2\gamma^{[a}{\rho_A}^{|b]c}
-\gamma^c{\rho_A}^{|ab})+4G^{-ab}\epsilon^{AB}\bar\psi_A\wedge c_B+
\nonumber\\
&+&4G^{+ab}\epsilon_{AB}\bar\psi^A\wedge c^B
+{i\over 4}\varepsilon^{abcd}(\bar\psi_A\wedge\gamma_c c^B
+\bar c_A\wedge\gamma_c\psi^B)
(2\bar\lambda_{iB}\gamma_d\lambda^{iA}-
\delta^A_B\bar\lambda_{iC}\gamma_d\lambda^{iC}),\nonumber\\
s\varepsilon^{ab}&=&{\varepsilon^a}_c\wedge\varepsilon^{cb}
+{R^{ab}}_{cd}\varepsilon^c \varepsilon^d-i\bar c_A(2\gamma^{[a}\rho^{A|b]c}
-\gamma^c\rho^{A|ab})\varepsilon_c+\nonumber\\
&-&i\bar c^A(2\gamma^{[a}{\rho_A}^{|b]c}
-\gamma^c{\rho_A}^{|ab})\varepsilon_c+
2G^{-ab}\epsilon^{AB}\bar c_A\wedge c_B+
\nonumber\\
&+&2G^{+ab}\epsilon_{AB}\bar c^A\wedge c^B
+{i\over 4}\varepsilon^{abcd}\bar c_A\wedge\gamma_c c^B
(2\bar\lambda_{iB}\gamma_d\lambda^{iA}-
\delta^A_B\bar\lambda_{iC}\gamma_d\lambda^{iC}),\nonumber\\
s\psi_A&=&-{\cal D}c_A+{1\over 4}\varepsilon^{ab}\gamma_{ab}\wedge\psi_A
-{i\over 2}Q\wedge c_A-{i\over 2}Q_{(0,1)}\wedge\psi_A+
2\rho_{A|ab}V^a\wedge \varepsilon^b+\nonumber\\
&-&2i \epsilon_{AB}G^+_{ab}\gamma^a (c^B\wedge V^b
+\psi^B\wedge \varepsilon^b)
+{i\over 4}(c_B V_a+\psi^B \varepsilon_a)
\bar\lambda^{iB}\gamma^a\lambda_{iA}+\nonumber\\
&+&{i\over 8}\gamma_{ab}(c_B V^b+\psi_B \varepsilon^b)
\left(2\bar\lambda^{iB}\gamma^a\lambda_{iA}-
\delta^B_A\bar
\lambda^{iC}\gamma^a\lambda_{iC}\right),\nonumber\\
sc_A&=&{1\over 4}\varepsilon^{ab}\gamma_{ab}\wedge c_A
-{i\over 2}Q_{(0,1)}\wedge c_A+\rho_{A|ab}\varepsilon^a\wedge
\varepsilon^b-2i \epsilon_{AB}G^+_{ab}\gamma^a c^B
\wedge \varepsilon^b+\nonumber\\
&+&{i\over 4}c_B\bar
\lambda^{iB}\gamma^a\lambda_{iA}\wedge \varepsilon_a+
{i\over 8}\gamma_{ab}c_B\left(2\bar
\lambda^{iB}\gamma^a\lambda_{iA}-\delta^B_A\bar
\lambda^{iC}\gamma^a\lambda_{iC}\right)\wedge \varepsilon^b,\nonumber\\
s\psi^A&=&-{\cal D}c^A+{1\over 4}\varepsilon^{ab}\gamma_{ab}\wedge\psi^A
+{i\over 2}Q\wedge c^A+{i\over 2}Q_{(0,1)}\wedge\psi^A+
2\rho^A_{|ab}V^a\wedge \varepsilon^b+
\nonumber\\
&-&2i \epsilon^{AB}G^-_{ab}\gamma^a (c_B\wedge V^b
+\psi_B\wedge \varepsilon^b)
+{i\over 4}(c^B V_a+\psi^B \varepsilon_a)
\bar\lambda_{iB}\gamma^a\lambda^{iA}+\nonumber\\
&+&{i\over 8}\gamma_{ab}(c^B V^b+\psi^B \varepsilon^b)
\left(2\bar\lambda_{iB}\gamma^a\lambda^{iA}-
\delta_B^A\bar
\lambda_{iC}\gamma^a\lambda^{iC}\right),\nonumber\\
sc^A&=&{1\over 4}\varepsilon^{ab}\gamma_{ab}\wedge c^A
+{i\over 2}Q_{(0,1)}\wedge c^A+\rho^A_{|ab}\varepsilon^a\wedge
\varepsilon^b-2i \epsilon^{AB}G^-_{ab}\gamma^a c_B
\wedge \varepsilon^b+\nonumber\\
&+&{i\over 4}c^B\bar
\lambda_{iB}\gamma^a\lambda^{iA}\wedge \varepsilon_a+
{i\over 8}\gamma_{ab}c^B\left(2\bar
\lambda_{iB}\gamma^a\lambda^{iA}-\delta_B^A\bar
\lambda_{iC}\gamma^a\lambda^{iC}\right)\wedge \varepsilon^b,\nonumber\\
sA_\Lambda&=&-dc_\Lambda-2{f_\Lambda}^{\Omega\Delta}A_\Omega\wedge c_\Delta
-2\epsilon_{AB}L_\Lambda \bar\psi^A\wedge c^B
-2\epsilon^{AB}\bar L_\Lambda
\bar\psi_A\wedge c_B+\nonumber\\
&+&2F_\Lambda^{ab}V_a\wedge \varepsilon_b
+i(f_\Lambda^i\bar\lambda^A_i
\gamma^a c^B\epsilon_{AB}+\bar f_{\Lambda i}\bar\lambda_A^i
\gamma^a c_B\epsilon^{AB})\wedge V_a+\nonumber\\
&+&i(f_\Lambda^i\bar\lambda^A_i
\gamma^a\psi^B\epsilon_{AB}+\bar f_{\Lambda i}\bar\lambda_A^i
\gamma^a\psi_B\epsilon^{AB})\wedge \varepsilon_a,\nonumber\\
sc_\Lambda&=&-{f_\Lambda}^{\Omega\Delta}c_\Omega\wedge c_\Delta-
\epsilon_{AB}L_\Lambda \bar c^A\wedge c^B-\epsilon^{AB}\bar L_\Lambda
\bar c_A\wedge c_B+F_\Lambda^{ab}\varepsilon_a\wedge \varepsilon_b+\nonumber\\
&+&i(f_\Lambda^i\bar\lambda^A_i
\gamma^a c^B\epsilon_{AB}+\bar f_{\Lambda i}\bar\lambda_A^i
\gamma^a c_B\epsilon^{AB})\wedge \varepsilon_a,\nonumber\\
s\lambda_{iA}&=&{1\over 4}\varepsilon^{ab}\wedge\gamma_{ab}
\lambda_{iA}-{i\over 2}\left(1+{2\over n}\right)Q_{(0,1)}\wedge
\lambda_{iA}-{{Q_{(0,1)}}_i}^j\lambda_{jA}+
\nabla_a\lambda_{iA}\varepsilon^a+\nonumber\\
&+&iP_{i|a}\gamma^a c^B
\epsilon_{AB}+G^{+ab}_i\gamma_{ab}c_A
+gC_i c_A,\nonumber\\
s\lambda^{iA}&=&{1\over 4}\varepsilon^{ab}\wedge\gamma_{ab}
\lambda^{iA}+{i\over 2}\left(1+{2\over n}\right)Q_{(0,1)}\wedge \lambda^{iA}
-{{Q_{(0,1)}}^i}_j\lambda^{jA}+\nabla_a\lambda^{iA}\varepsilon^a+\nonumber\\
&+&iP^i_{|a}\gamma^a c_B
\epsilon^{AB}+G_{ab}^{-i}\gamma^{ab}c^A
+gC^i c^A,\nonumber\\
s z_i&=&-gc_\Lambda k^\Lambda_i(z)+Z_{i|a}\varepsilon^a+\bar\lambda_i^A
c_A,\nonumber\\
s \bar z^i&=&-gc_\Lambda k^{i\Lambda}(\bar z)+\bar Z^i_{|a} \varepsilon^a
+\bar\lambda^i_A c^A.
\label{brstalgebra}
\end{eqnarray}
In Eq.\ (\ref{brstalgebra})
$Q_{(0,1)}$ and ${{Q_{(0,1)}}_i}^j$ are obtained by the one-forms
$Q$ and ${Q_i}^j$
upon substitution of $\nabla z_i$ with
$Z_{i|a}\varepsilon^a+\bar\lambda_i^A c_A$,
$\nabla z_i$ with $\bar Z^i_{|a} \varepsilon^a
+\bar\lambda^i_A c^A$ and of $A_\Lambda$ with $c_\Lambda$.
In particular, $Q_{(0,1)}=-{i\over 2}(G^i Z_{i|a}-G_i\bar
Z^i_{|a})\varepsilon^a
-{i\over 2}(G^i\bar\lambda_i^A c_A-G_i\bar \lambda^i_A c^A)$.

The BRST algebra of the twisted theory is the above algebra when one
implements the topological twist and the topological shift,
as explained in Ref.\ \cite{anselmifre}. From now on, when we shall refer to
the above algebra, this implementation will be understood.
The explicit twist is realized as follows
\begin{equation}
\matrix{
\psi_{\alpha A}\rightarrow \psi_{\alpha \dot A}, &
\psi^{\dot \alpha A}\rightarrow \psi^{\dot\alpha \dot A},\cr
\lambda_{i\alpha A}\rightarrow \lambda_{i\alpha\dot A},&
\lambda^{i\dot\alpha A}\rightarrow \lambda^{i\dot\alpha\dot A},\cr
\epsilon_{AB}\rightarrow \epsilon_{\dot A \dot B},&
\epsilon^{AB}\rightarrow -\epsilon^{\dot A \dot B},}
\label{explicittwist}
\end{equation}
while the topological shift is obtained by
\begin{equation}
c^{\dot\alpha\dot A}\rightarrow -{i\over 2}e\varepsilon^{\dot \alpha\dot A}
+c^{\dot\alpha\dot A}.
\end{equation}
Here, $e$ is an object that rearranges the form-number, ghost-number
and statistics in the correct way and that appears only in
the intermediate steps of the twist. It will be called
the {\sl broker}.
The broker
is a zero-form with fermionic statistics and ghost number one.
$e^2$ has even ghost number and Bose statistics, hance it
can be set equal to a number and in our
notation we normalize it as $e^2=1$.

We now rewrite the most relevant twisted-shifted BRST transformations
up to nonlinear terms. To this purpose, note that, when $z_i$ and $\bar z^i$
tend to zero, then
$a\rightarrow 1$; $L_0,\bar L_0\rightarrow 1$;
${f_i}^j\rightarrow \delta_i^j$;
${(g^{1\over 2})_i}^j\rightarrow \delta_i^j$; $G^+_{ab}\rightarrow
-{1\over 2}F^{+0}_{ab}$;
$G^{i-}_{ab}\rightarrow -{1\over 2}F^{i-}_{ab}$. Let us define
(note that the gauginos are expressed in the N=3 notation, namely
$\lambda_{iA}$, $\lambda^{iA}$)
\begin{equation}
\matrix{
\tilde\psi^a={e\over 2}\psi_{\alpha \dot A}(\bar\sigma^a)^{\dot A
\alpha}, & \tilde\psi^{ab}=-e{(\bar \sigma^{ab})^{\dot A}}_{\dot \alpha}
{\psi^{\dot \alpha}}_{\dot A},&\tilde \psi=-e{\psi_{\dot \alpha}}^{\dot A}
\delta_{\dot A}^{\dot \alpha},\cr
C^a={e\over 2}c_{\alpha \dot A}(\bar\sigma^a)^{\dot A
\alpha}, & C^{ab}=-e{(\bar \sigma^{ab})^{\dot A}}_{\dot \alpha}
{c^{\dot \alpha}}_{\dot A},& C=-e{c_{\dot \alpha}}^{\dot A}
\delta_{\dot A}^{\dot \alpha},\cr
\lambda_i={e\over 2}{\lambda_i}_{\alpha \dot A}(\bar\sigma^a)^{\dot A
\alpha}V_a, & {\lambda^i}^{ab}=-e{(\bar \sigma^{ab})^{\dot A}}_{\dot \alpha}
{{\lambda^i}^{\dot \alpha}}_{\dot A},& \tilde\lambda^i=-e
{{\lambda^i}_{\dot \alpha}}^{\dot A}\delta_{\dot A}^{\dot \alpha}.}
\label{definit}
\end{equation}
As an example of the action of the broker $e$, note that,
while ${1\over 2}\psi_{\alpha \dot A}(\bar\sigma^a)^{\dot A
\alpha}$ is a one-form, is a fermion
and has ghost number zero,
the true topological ghost
$\tilde\psi^a$  must be a one-form, with ghost number one and it is a boson.
In Ref.\ \cite{anselmifre} the broker was not explicitly introduced,
although it was implicitly assumed.

Up to nonlinear terms, we obtain
\begin{equation}
\matrix{
s V^a=\tilde\psi^a-d\varepsilon^a+\varepsilon^{ab}\wedge V_b,&
s \varepsilon^{a}=C^a,\cr
s \varepsilon^{ab}=-{1\over 2}{F_0^+}^{ab},&
s\tilde\psi^a=-d C^a+{1\over 2}{F^+}^{ab}_0\wedge V_b,\cr
s\tilde\psi^{ab}=-dC^{ab}+{i\over 2}{\omega^-}^{ab},&
s \tilde\psi =-dC ,\cr
s C^{a}=0,&
s C^{ab}={i\over 2}{\varepsilon^-}^{ab},\cr
s C=0,&
s\lambda_i={1\over 2}dz_i,\cr
s{\lambda^i}^{ab}=i F_i^{-ab},&
s\tilde\lambda^i=0,\cr
sA_i=-dc_i+\lambda_i,&
sc_i=-{1\over 2}z_i,\cr
sz_i=0,&
s\bar z^i={i\over 2}\tilde \lambda^i,\cr
s A_0= i \tilde\psi -dc_0,&
s c_0 =-{1\over 2}+iC.}
\label{simply}
\end{equation}
Here $F^{-ab}={1\over 2}(F^{ab}+{i\over 2}\varepsilon^{abcd}F_{cd})$
(with respect to Ref.\ \cite{anselmifre} there is, in particular,
a sign difference
in the conventions for $\gamma_5$ and $\varepsilon_{abcd}$).

{}From Eq.s (\ref{definit}) and (\ref{simply})
we can directly identify what are the topological ghosts,
the topological antighosts (up to
interaction terms) and the topological gauge-fixings.
More generally, one retrieves the topological meaning of the twisted versions
of all the fields of the original theory.
$\tilde\psi^a$ are the topological ghosts associated to the graviton,
$\lambda_i$ those associated to the matter vectors;
the corresponding topological antighosts are $\tilde\psi^{ab}$
and $\lambda^{iab}$, respectively.
The ghosts for ghosts are
$C^a$, $F^{+ab}_0$ and $z_i$, respectively
for diffeomorphisms, Lorentz rotations and
gauge transformations.
$\bar z^i$ are antighosts for ghosts, while
$C^{ab}$ and $C$ are extraghosts.
Let us discuss the gauge-fixings.
They involve complicated expressions depending on the various fields
(even in the
topological $\sigma$-model in two dimensions \cite{witten2}
one finds convenient to impose a topological gauge-fixing depending
on the ghosts),
but they can be equivalently read when all the ghosts are set to zero,
because in the
minimum of the BRST action all the ghosts are zero by definition.
To this purpose, the interaction terms are negligible (they
always contain ghosts). Our expectations are confirmed: the theory
does indeed describe
Yang-Mills instantons $F_i^{-ab}=0$ in a background gravitational
instanton $\omega^{-ab}=0$ (the Wick rotation to the Euclidean is
of course understood, as in Ref.\ \cite{anselmifre})\footnotemark
\footnotetext{Note that the BRST variation of the topological gravitational
antighost $\tilde\psi^{ab}$ contains, in addition to the gauge-fixing
$\omega^{-ab}$, also the derivative of the extraghost $C^{ab}$.
As explained in Ref.\ \cite{anselmifre}, this is due to the redundancy of
the gauge conditions $\omega^{-ab}=0$.}.

We note that there are more observables than those we have constructed
by means of the minimal BRST algebra (\ref{gaugefree}). They involve
also antighosts. In fact there is
another noticeable
differential form which is closed but not exact and which could be a source
of nontrivial observables, namely the K\"ahler form $K$. In fact the K\"ahler
potential $G$ exists only locally and $K=dQ$ is only a local statement.
The associated descent equations still give observables, however so far
we have not revealed their deep meaning (if any). The K\"ahler form
and its extended version are constructed with both
ghosts and antighosts, while one usually uses only ghosts.
We must remark that the topological Yang-Mills theory we have found
is {\sl not} exactly Witten's topological Yang-Mills theory coupled to gravity.
In fact, Witten's theory is described by a flat K\"ahler manifold
(and $Q$ exists globally, so $K$ is not interesting), while our theory
corresponds to ${SU(1,n)\over SU(n)\otimes U(1)}$ and $K$ cannot be
globally exact \cite{dauriaferrarafre}, so it cannot be
{\sl a priori} discarded. One has
\begin{equation}
K=i{g_i}^j\nabla z_j\wedge \nabla \bar z^i+
{i\over 2}g(G_i k^{i\Lambda}-G^i k_i^\Lambda)
(dA_\Lambda+{f_\Lambda}^{\Sigma\Pi}A_\Sigma\wedge A_\Pi).
\end{equation}
The descent equations derived from $\hat d \hat K=0$ give
the following observables
\begin{eqnarray}
{\cal O}^{(0)}&=&K_{(0,2)},\nonumber\\
{\cal O}^{(1)}_\gamma&=&\int_\gamma K_{(1,1)},\nonumber\\
{\cal O}^{(2)}_S&=&\int_S K,
\end{eqnarray}
where $\gamma$ and $S$ are one- and two-dimensional cycles, while
\begin{eqnarray}
K_{(0,2)}&=&i{g_i}^j(Z_{j|a}\varepsilon^a+\bar\lambda_j^A c_A)
\wedge (\bar Z^i_{|a} \varepsilon^a
+\bar\lambda^i_B c^B)+\nonumber\\
&-&{i\over 2}g(G_j k^{j\Lambda}-G^j k_j^\Lambda)
(\epsilon_{AB}L_\Lambda \bar c^A\wedge c^B+\epsilon^{AB}\bar L_\Lambda
\bar c_A\wedge c_B-F_\Lambda^{ab}\varepsilon_a\wedge \varepsilon_b+\nonumber\\
&-&i(f_\Lambda^i\bar\lambda^A_i
\gamma^a c^B\epsilon_{AB}+\bar f_{\Lambda i}\bar\lambda_A^i
\gamma^a c_B\epsilon^{AB})\wedge \varepsilon_a),\nonumber\\
K_{(1,1)}&=&i{g_i}^j(Z_{j|a}\varepsilon^a+\bar\lambda_j^A c_A)
\wedge \nabla \bar z^i+i{g_i}^j\nabla z_j
(\bar Z^i_{|a} \varepsilon^a
+\bar\lambda^i_A c^A)+
\nonumber\\
&-&{i\over 2}g(G_j k^{j\Lambda}-G^j k_j^\Lambda)
(2\epsilon_{AB}L_\Lambda \bar\psi^A\wedge c^B
+2\epsilon^{AB}\bar L_\Lambda
\bar\psi_A\wedge c_B+\nonumber\\
&-&2F_\Lambda^{ab}V_a\wedge \varepsilon_b
-i(f_\Lambda^i\bar\lambda^A_i
\gamma^a c^B\epsilon_{AB}+\bar f_{\Lambda i}\bar\lambda_A^i
\gamma^a c_B\epsilon^{AB})\wedge V_a+\nonumber\\
&-&i(f_\Lambda^i\bar\lambda^A_i
\gamma^a\psi^B\epsilon_{AB}+\bar f_{\Lambda i}\bar\lambda_A^i
\gamma^a\psi_B\epsilon^{AB})\wedge \varepsilon_a).
\end{eqnarray}

The correspondence between the gauge-free algebra (\ref{gaugefree})
and the complete BRST algebra (\ref{brstalgebra}) is realized by the
following identifications
\begin{eqnarray}
\psi^a&=&i(\bar c_A\wedge \gamma^a\psi^A+
\bar \psi_A\wedge \gamma^a c^A)-A^{ab}\wedge \varepsilon^b
=\tilde\psi^a+\cdots,\nonumber\\
\phi^a&=&i\bar c_A\wedge\gamma^a c^A=C^a+\cdots,\nonumber\\
\chi^{ab}&=&sA^{ab}-A^{ac}{\varepsilon_c}^b+\varepsilon^{ac}
{A_c}^b+2{R^{ab}}_{cd}V^c\varepsilon^d+i(\varepsilon_c\bar\psi_A+V_c \bar c_A)
(2\gamma^{[a}\rho^{A|b]c}
-\gamma^c\rho^{A|ab})+\nonumber\\
&+&i(\varepsilon_c \bar\psi^A+ V_c \bar c^A)
(2\gamma^{[a}{\rho_A}^{|b]c}
-\gamma^c{\rho_A}^{|ab})+4G^{-ab}\epsilon^{AB}\bar\psi_A\wedge c_B+
\nonumber\\
&+&4G^{+ab}\epsilon_{AB}\bar\psi^A\wedge c^B
+{i\over 4}\varepsilon^{abcd}(\bar\psi_A\wedge\gamma_c c^B
+\bar c_A\wedge\gamma_c\psi^B)
(2\bar\lambda_{iB}\gamma_d\lambda^{iA}-
\delta^A_B\bar\lambda_{iC}\gamma_d\lambda^{iC}),\nonumber\\
\psi_i&=&2\epsilon_{AB}L_i \bar\psi^A\wedge c^B
+2\epsilon^{AB}\bar L_i
\bar\psi_A\wedge c_B+\nonumber\\
&-&2F_i^{ab}V_a\wedge \varepsilon_b
-i(f_i^j\bar\lambda^A_j
\gamma^a c^B\epsilon_{AB}+\bar f_{i j}\bar\lambda_A^j
\gamma^a c_B\epsilon^{AB})\wedge V_a+\nonumber\\
&-&i(f_i^j\bar\lambda^A_j
\gamma^a\psi^B\epsilon_{AB}+\bar f_{i j}\bar\lambda_A^j
\gamma^a\psi_B\epsilon^{AB})\wedge \varepsilon_a
=-\lambda_i+\cdots,\nonumber\\
\phi_i&=&-\epsilon_{AB}L_i \bar c^A\wedge c^B
-\epsilon^{AB}\bar L_i
\bar c_A\wedge c_B+F_i^{ab}\varepsilon_a\wedge \varepsilon_b+\nonumber\\
&+&i(f_i^j\bar\lambda^A_j
\gamma^a c^B\epsilon_{AB}+\bar f_{i j}\bar\lambda_A^j
\gamma^a c_B\epsilon^{AB})\wedge \varepsilon_a
=-{1\over 2}z_i+\cdots,\nonumber\\
\eta^{ab}&=&{R^{ab}}_{cd}\varepsilon^c \varepsilon^d-i\bar
c_A(2\gamma^{[a}\rho^{A|b]c}
-\gamma^c\rho^{A|ab})\varepsilon_c
-i\bar c^A(2\gamma^{[a}{\rho_A}^{|b]c}
-\gamma^c{\rho_A}^{|ab})\varepsilon_c+\nonumber\\&+&
2G^{-ab}\epsilon^{AB}\bar c_A\wedge c_B+
2G^{+ab}\epsilon_{AB}\bar c^A\wedge c^B+\nonumber\\
&+&{i\over 4}\varepsilon^{abcd}\bar c_A\wedge\gamma_c c^B
(2\bar\lambda_{iB}\gamma_d\lambda^{iA}-
\delta^A_B\bar\lambda_{iC}\gamma_d\lambda^{iC})=-{1\over 2}F^{+ab}_0+\cdots,
\label{matching}
\end{eqnarray}
where $A^{ab}\wedge V_b=i\bar \psi_A\wedge\gamma^a\psi^A$
and the dots stand for nonlinear corrections.

Now we write the gauge fermion $\Psi$, the BRST variation of which is the
quadratic part of the N=2 lagrangian, after topological
twist and topological shift.
\begin{eqnarray}
\Psi&=&-16i(B^{ab}-i\omega^{-ab}+2dC^{ab})\wedge \tilde\psi_{ac}
\wedge V_b\wedge V^c+8iF_0\wedge \psi^a\wedge V_a+\nonumber\\
&+&\left({2\over 3}\eta^{ab}\varepsilon_{ab}
-{1\over 6}(M_{iab}-2iF^{-}_{iab})\lambda^{iab}\right)
\varepsilon_{cdef}V^c\wedge V^d\wedge V^e\wedge V^f+\nonumber\\
&+&{4\over 3}\lambda_i^a d\bar z^i\wedge\varepsilon_{abcd}V^b\wedge V^c\wedge
V^d.
\end{eqnarray}
Here,
$B^{ab}$ and $M^{iab}$ are Lagrange multipliers ($s\tilde\psi^{ab}=B^{ab}$,
$s\lambda^{iab}=M^{iab}$, $sB^{ab}=0$, $sM^{iab}=0$), while
$\lambda_i^a$ is such that $\lambda_i=\lambda_i^a V_a$.

\section{The general structure of
the twisting procedure and quaternionic topological $\sigma$-model}
\label{sectquater}

In this section we discuss the topological twist of quaternionic matter
multiplets \cite{dauriaferrarafre} coupled to N=2 supergravity.
We shall not develop the entire formalism in full detail, living it for a
future publication, but we shall concentrate on some of its relevant aspects.

We already anticipated in the introduction that the twisting procedure
as described by Witten \cite{witten} needs some modifications in order to
work correctly. First of all, as shown in Ref.\ \cite{anselmifre},
the twist acts on the Lorentz group and does not touch the space-time indices.
This was straightforward in the case of pure supergravity, since
all the fields are one-forms, i.\ e.\ they are all on the same footing
as far as space-time indices are concerned. Consequently the twist on
the Lorentz group works in exactly the same way as the twist
described by Witten. However, when studying the
case of the Yang-Mills theory, one has to face the problem that the vector
bosons $A_i$ are one forms and Lorentz scalars, while the gauginos
$\lambda_i^A$ and $\lambda^i_A$ are zero-forms and Lorentz spinors. If you
are in flat space, you can mix Lorentz and Einstein indices and so the
twist can work in the way described by Witten. However, Witten
himself notes \cite{witten} that his method works only in flat
space, even if the result is valid in any curved space. If we
follow our method, this problem is simply absent. We remain
in the most general curved space and act only on the Lorentz indices.
At this point, the twisted vector boson is still a one-form and a Lorentz
scalar, while the twisted left handed gaugino $\lambda_i^a\equiv
{e\over 2}{\lambda_i}_{\alpha \dot A}(\bar\sigma^a)^{\dot A
\alpha}$ is a zero-form and a Lorentz vector.
{}From (\ref{simply}) you immediately read that the true topological antighost
is not simply $\lambda_i^a$, but $\lambda_i=
\lambda_i^aV_a$, i.\ e.\ the object that you obtain from the simple
twist ($\lambda_i^a$) must be contracted with the vierbein $V^a$.
$\lambda_i$ is a one-form and a Lorentz scalar, as desired.

In order to show that the contraction with a vielbein
plays a substantial role in the twisting procedure,
one would like to exhibit a case in which
this step is so important that no result can be
obtained without it (even in flat space). This
is precisely the case of the quaternionic $\sigma$-model. The multiplet
consists of $(q^i,\zeta_I,\zeta^I)$, where $\zeta_I$ and
$\zeta^I$ are the left
handed and right handed components of the spinors ($I=1,\ldots 2m$),
while $q^i$ are the
coordinates of a 4$m$-dimensional manifold ${\cal Q}(m)$
($i=1,\ldots 4m$),
with a quaternionic structure, namely
possessing three complex structures
$J^x$, $x=1,2,3$, fulfilling the quaternionic
algebra. Specifically ${\cal Q}(m)$ is a Hyperk\"ahler manifold when gravity
is not dynamical (i.e.\
it is external), while it is a quaternionic manifold
when gravity is dynamical.
As you see, no field has indices of $SU(2)_I$, i.\ e.\
all the fields are singlets under the internal $SU(2)$.
Consequently, the usual twisting procedure acts trivially on hypermultiplets:
the Lorentz scalars remain Lorentz scalars
and the spinors remain spinors. In a moment we shall show
how this problem can be solved by means of a contraction with
a suitable vielbein.

The general feature of ${\cal Q}(m)$ is that
its holonomy group $Hol({\cal Q}(m))$ is contained in $SU(2)\otimes Sp(2m)$.
This $SU(2)$ is nothing but $SU(2)_I$ \cite{dauriaferrarafre}.
In the Hyperk\"ahler case, the $SU(2)$ part of the spin connection
of ${\cal Q}(m)$ is flat, while in the quaternionic case
its curvature is proportional to $\Omega^x=h_{ik}(J^x)^k_jdq^i\wedge dq^j$,
where $h_{ij}$ is the metric of ${\cal Q}(m)$. In both
cases one can exploit another $SU(2)$, which will be
denoted by $SU(2)_Q$, namely the $SU(2)$ factor in
the $SU(2)\otimes SO(m)$ maximal subgroup of $Sp(2m)$. We shall see that
the twisting procedure requires also a redefinition of
$SU(2)_L$, namely
\begin{equation}
SU(2)_L\rightarrow SU(2)_L^\prime={\rm diag}[SU(2)_L\otimes SU(2)_Q].
\end{equation}
Summarizing, the complete twisting procedure can be divided in the
following three steps. Step A corresponds to the redefinitions of $SU(2)_L$,
$SU(2)_R$
and ghost number $U(1)_g$
\begin{eqnarray}
SU(2)_L&\rightarrow & SU(2)_L^\prime={\rm diag}
[SU(2)_L\otimes SU(2)_Q],\nonumber\\
SU(2)_R&\rightarrow & SU(2)_R^\prime={\rm diag}
[SU(2)_R\otimes SU(2)_I],\nonumber\\
U(1)_g&\rightarrow & U(1)_g^\prime={\rm diag}
[U(1)_g\otimes U(1)_I],\nonumber\\
^c(L,R,I,Q)^g_f &\rightarrow & (L\otimes Q,R\otimes I)^{g+c}_f,
\end{eqnarray}
where $Q$ denotes the representation of $SU(2)_Q$.
Step B is the correct identification
of the topological ghosts
(fields with $g+c=1$ from $g=0$, $c=1$) by contraction
with a suitable vielbein (if it exists). Step 3 is the topological shift,
namely the shift by a constant
of the $(0,0)^0_0$ field coming from the right handed components
of the supersymmetry ghosts.

Let us see how the contraction with a suitable vielbein can help
when the usual twisting procedure does not give directly
the true topological ghosts (i.e.\ it gives objects with the
wrong spin assignment).
Since the hypermultiplets are made of zero-forms,
the vierbein $V^a$ cannot help us. Fortunately, however, there
{\sl is } a vielbein that does the job, namely the quaternionic vielbein
${\cal U}^{A I}_i$ ($A=1,2$ is an index of $SU(2)_I$)
\cite{dauriaferrarafre}.
We can for example take the contraction ${\cal U}^i_{A I}\bar c^A
\zeta^I$, where ${\cal U}^i_{A I}$ is the inverse vielbein.
After the topological shift, this expression becomes
$-{i\over 2}e{\cal U}^i_{\dot A I}\zeta^{\dot A I}$, up
to interaction terms, and is the natural candidate to become the topological
ghost (it is also the {\sl only} candidate).
Here is another novelty: the topological ghost is constructed with
the {\sl right} handed components of the fermions, {\sl not}
the left handed ones. This means that the R-duality charge of
$\zeta^I$ is $+1$ and that of $\zeta_I$ is $-1$, the opposite
of what happens in the other cases that we have studied. This is not
completely surprising, because the reasoning of Section \ref{general}
that established the R-duality charges of gravitinos and gauginos
was essentially based on the effects of the usual redefinition
of $SU(2)_R$ on the
representations of the Lorentz group, effects that are absent
in the present case. From Ref.\ \cite{dauriaferrarafre} one can convince
oneself
that this is in fact the correct charge assignment.

We report here only those terms in the BRST variations of the fields that
correspond to supersymmetries, in order to spot the nature of
the instantons described by the theory,
reading the topological gauge-fixings.
\begin{eqnarray}
\delta q^i&=&{\cal U}^i_{A I}(\epsilon^{AB}C^{IJ}
\bar c_B\zeta_J+\bar c^A\zeta^I),\nonumber\\
\delta\zeta_I&=&i{\cal U}_a^{BJ}\gamma^a c^A\epsilon_{AB}C_{IJ},\nonumber\\
\delta\zeta^I&=&i{\cal U}_a^{AI}\gamma^a c_A,
\end{eqnarray}
where $C_{IJ}$ is the flat $Sp(2m)$ invariant metric
while ${\cal U}_a^{AI}$ is the supercovariantized
derivative of the quaternionic field $q^i$ with indices flattened both with
respect to spacetime and with respect to the quaternionic manifold
via the corresponding vielbeins.
\begin{equation}
{\cal U}_a^{A I}=V^\mu_a({\cal U}_i^{A I}\partial_\mu q^i-
\epsilon^{AB}C^{IJ}
\bar{\psi_\mu}_B\zeta_J-\bar {\psi_\mu}^A\zeta^I).
\end{equation}
The topological shift gives, up to nonlinear terms,
\begin{eqnarray}
\delta q^i&=&-{i\over 2}e{\cal U}^i_{\dot A I}(\zeta^{\dot A})^I
\equiv\xi^i,\nonumber\\
\delta(\zeta_\alpha)_I&=&{e\over 2}{\cal U}_a^{\dot B J}
(\sigma^a)_{\alpha \dot B}C_{IJ},\nonumber\\
\delta(\zeta^{\dot A})^I&=&0,
\end{eqnarray}
{}From this equation we realize that the topological symmetry is indeed
the expected one for a $\sigma$-model, namely
the map $q^i:M_{spacetime}\rightarrow {\cal Q}(m)$ can be continuously
deformed.
The topological ghosts $\xi^i$ are exactly what we anticipated.
In order to correctly identify the topological antighosts, we have to write
the index $I$ as the pair $(\alpha,k)$, where $\alpha=1,2$ is the doublet
index of $SU(2)_Q$ and $k=1,\ldots m$ is the
vector index of $SO(m)$, such that
$C_{IJ}=C_{(\alpha,k)(\beta,l)}$
takes the form $\epsilon_{\alpha\beta}\delta_{kl}$.
Now we write
\begin{equation}
\delta(\zeta_\alpha)_{\beta k}={e\over 2}{\cal U}_a^{\dot B \gamma l}
(\sigma^a)_{\alpha \dot B}\epsilon_{\beta\gamma}\delta_{kl}=
{e\over 2}{\cal U}_a^{\dot B \gamma k}
(\sigma^a)_{\alpha \dot B}\epsilon_{\beta\gamma}.
\end{equation}
At this point we can introduce the vielbein $E_i^{ak}\equiv {1\over 2}
{\cal U}^{\dot A
\alpha k}_i(\sigma^a)_{\alpha \dot A}$ and the true topological antighosts
$\zeta^{+ab}_k=-e{(\sigma^{ab})_\alpha}^\beta\epsilon^{\alpha\gamma}(\zeta_\beta)
_{\gamma k}$ and $\zeta_k=-e\epsilon^{\alpha\beta}(\zeta_\alpha)_{\beta k}$,
which, under the Lorentz group transform
as $(1,0)$ and $(0,0)$ respectively. One finds
\begin{eqnarray}
\delta \zeta^{+ab k}&=&2V^{\mu[a}E_i^{b]^+k}\partial_\mu q^i,
\nonumber\\
\delta\zeta^k&=&V^\mu_a E^{ak}_i\partial_\mu q^i,
\end{eqnarray}
where $[ab]^+$ means selfdualization in the indices $a,b$. Thus we see
that {\sl both} $\zeta^{+ab}_k$ and $\zeta_k$ are topological antighosts
(otherwise we would have not enough equations to fix the gauge completely).
In the previously studied cases, instead, the $(0,1)$ components
were the only topological antighosts, while
the $(0,0)$ component permitted to
fix the gauge freedom of the topological ghosts (directly related  to
the gauge freedom of the gauge freedom, which now is missing).
Thus, the instantons described by this
theory (which we name {\sl hyperinstantons}) are
given by the following equations
\begin{eqnarray}
V^{\mu[a}E_i^{b]^+k}\partial_\mu q^i &=&0,
\nonumber\\
V^\mu_a E^{ak}_i \partial_\mu q^i &=&0.
\label{inst1}
\end{eqnarray}
In a certain sense, Eq.\ (\ref{inst1}) define a condition of holomorphicity
of the maps $M_{spacetime}\rightarrow {\cal Q}(m)$ with respect to
the three complex (or almost complex) structures $J^x$ of
${\cal Q}(m)$. For this reason we find it proper to name
triholomorphic a map $q$ satisfying Eq.\ (\ref{inst1}). In conclusion,
in the same way as the instantons of topological $\sigma$-models in D=2
are given by holomorphic maps, those of topological $\sigma$-models
in D=4 are given by triholomorphic maps.

If gravity is external (${\cal Q}(m)$ is Hyperk\"ahler) then the gravitational
background should be restricted by the need to have N=2 global supersymmetry,
however, the proof that the solutions to the above equations are indeed
instantons works for any background and is based on the following identity
\begin{eqnarray}
\int_{\cal M} &d^4x\sqrt{-g}& g^{\mu\nu}\partial_\mu q^i\partial_\nu q^j
h_{ij}=
2\int_{\cal M} d^4x
\sqrt{-g}[(V^\mu_a E^{ak}_i \partial_\mu q^i)^2+4(V^{\mu[a}E_i^{b]^+k}
\partial_\mu q^i)^2]+\nonumber\\
&-4i\int_{\cal M}&E^{[a k}\wedge E^{b]^- k}\wedge V_a \wedge V_b,
\label{dim1}
\end{eqnarray}
where $h_{ij}=2E_i^{ak}E_j^{bk}\eta_{ab}$ is the metric of ${\cal Q}(m)$,
while $E^{ak}\equiv E_i^{ak} dq^i$. The form
$E^{[a k}\wedge E^{b]^- k}\wedge V_a \wedge V_b$ is proportional to
$\Omega^x\wedge V_a \wedge V_b$ (the coefficient, which is a
numerical matrix $M_x^{ab}$ antiselfdual in $ab$ is not important),
where $\Omega^x$ are the 2-forms introduced above.
$\Omega_x$ are closed forms of ${\cal Q}(m)$ if ${\cal Q}(m)$ is
Hyperk\"ahler. Consequently, in such a case the last term of
(\ref{dim1}) is a topological invariant and this completes our proof.

In the case gravity is dynamical (${\cal Q}(m)$ is quaternionic)
there exist three forms $\omega_x$ such that
\begin{eqnarray}
d&\Omega_x&+\varepsilon_{xyz}\omega^y\wedge\Omega^z=0,\nonumber\\
d&\omega_x&+{1\over 2}\varepsilon_{xyz}\omega^y\wedge\omega^z=\Omega_x.
\end{eqnarray}
The definition of the curvatures changes drastically with respect
to the case of pure N=2 supergravity
\cite{dauriaferrarafre}, in the sense that the curvature
$\rho^A$ of the right handed components of the gravitinos contains
a term that modifies the gravitational topological gauge-fixing,
after performing the topological twist and the topological shift.
This means that the gravitational instantons
are no longer described by an antiselfdual spin connection.
As a matter of fact $\rho^A={\cal D}\psi^A+{i\over 2}\epsilon^{AB}\epsilon_{CD}
{(\sigma_x)_B}^C \psi^D$, where ${(\sigma_x)_A}^B$ are the Pauli matrices
and the resulting instantons are given by
\begin{equation}
\omega^{-ab}-{i\over 2}M_x^{ab}\omega^x=0.
\label{inst2}
\end{equation}
There exist only one matrix with the properties of $M^{ab}_x$, up
to a multiplicative constant, and this constant can be fixed by the fact
that $M_x^{ac}M_y^{db}\eta_{cd}{\varepsilon^{xy}}_z=2i M_x^{ab}$
(see for example section 5 of \cite{billo}).
The proof that the hyperinstantons that solve Eq.s (\ref{inst1})
and (\ref{inst2}) are effectively instantons follows from the fact that the
total kinetic
lagrangian (Einstein lagrangian plus $\sigma$-model kinetic lagrangian)
can be written as a sum of squares of the left hand sides of the above
equations up to a total derivative
\begin{eqnarray}
{\cal L}_{kin}&=&\varepsilon_{abcd}R^{ab}\wedge V^c\wedge V^d
-{1\over 6}\varepsilon_{abcd}V^a\wedge V^b\wedge V^c\wedge V^d
g^{\mu\nu}h_{ij}\partial_\mu q^i\partial_\nu q^j=\nonumber\\
&=&4i(\omega^{-ab}-{i\over 2}M^{ab}_x\omega^x)\wedge
(\omega_{-ac}-{i\over 2}{M_{ac}}_y\omega^y)\wedge V_b\wedge V^c+\nonumber\\
&-&{1\over 3}\varepsilon_{cdef}V^c\wedge V^d\wedge V^e\wedge V^f
[4(V^{\mu[a}E_i^{b]^+k}\partial_\mu q^i)^2+
(V^\mu_a E^{ak}_i \partial_\mu q^i)^2]+\nonumber\\
&+&{\rm total \hskip .1truecm derivative}.
\end{eqnarray}

As an example, let us consider the simplest case, namely the case
$m=1$, ${\cal Q}(1)=H^1$, with the standard flat metric. We have
${\cal U}^i_{\dot A\alpha}=(\sigma^i)_{\alpha \dot A}$
and $E^a_i=\delta^a_i$.

The hyperinstantons satisfy
\begin{eqnarray}
V^{\mu [a}\partial_\mu q^{b]^+} &=&0,
\nonumber\\
V^\mu_a \partial_\mu q^a &=&0.
\end{eqnarray}

If we further specialize the example, namely we choose flat spacetime metric,
we have
\begin{eqnarray}
\partial_{[\mu} q_{\nu]^+}&=&0,
\nonumber\\
\partial_\mu q^\mu &=&0.
\end{eqnarray}
If you imagine that $q_\mu$ is an abelian four vector,
the hyperinstantons are
the selfdual solutions in the Lorentz gauge. But now $\partial_\mu q^\mu=0$
is a true equation and not a choice of gauge. In particular, all harmonic
forms $q=q_\mu dx^\mu$ are solutions (they would be the residual gauge freedom
in the interpretation of $q_\mu$ as a four potential and so they would
not be true solutions).

\section{Conclusions}
\label{concl}

We have seen that, with appropriate procedure and relying on an appropriate
symmetry (R-duality), all N=2, D=4 theories can be topologically twisted,
just as it happens of N=2 theories in two dimensions. This possibility
introduces a set of new topological field theories, each of which describes
intersection theory in the moduli-space of certain interesting geometrical
structures. Some of these structures are, as far as we know, new or at
least not well estabilished in the mathematical literature.

To be specific, let us enumerate these theories.

i) The twist of N=2 $\sigma$-models in flat background, whose target space is a
Hyperk\"ahler manifold, introduces the notion of a topological
hyperk\"ahlerian $\sigma$-model, where the appropriate instantons are the
triholomorphic maps (hyperinstantons). Correlation functions in this theory
will be intersection integrals in the moduli-space of triholomorphic maps:
a subject that to our knowledge has not been so far developed and certainly
deserves careful investigation.

ii) The twist of N=2 supergravity minimally coupled to vector multiplets
yields a topological theory where the instantons are gauge instantons living
in the background of gravitational instantons. The moduli-space of these
structures is the arena
where correlation functions of our theory have to be calculated. Making
an analogy with the 2-dimensional world, our theory stands to
topological Yang-Mills theory as the
topological matter models coupled to topological gravity stand to
pure topological minimal models in D=2.

iii) Similarly, twisting N=2 $\sigma$-models coupled to N=2
supergravity, one obtains a topological $\sigma$-model where the target space
is
quaternionic and which interacts with topological gravity.
The instantons of this theory are interesting objects. They
correspond to the quaternionic analogue of triholomorphic maps
living in the background of generalized
gravitational instantons.
The space-time spin connection is no longer selfdual
but its antiselfdual part is identified with the $SU(2)$ part of the spin
connection
on the quaternionic manifold. This is a phenomenon similar to the embedding of
the
spin connection into the gauge connection occurring in string
compactifications.

iv) Twisting the complete N=2 matter coupled supergravity, one obtains
a topological theory where all the above instantons are fused together:
gravitational, gauge and hyperinstantons. To our knowledge, no study
of the moduli-space of such structures has been attempted.

v) Alternatively, one can also study the twist on N=2
hyperk\"ahlerian $\sigma$-models coupled to N=2 super Yang-Mills. In this
case we have the fusion of gauge and hyperinstantons.

\acknowledgments
Its a pleasure to thank
R.\ D'Auria, S.\ Ferrara and P.\ Soriani
for illuminating and inspiring discussions.

\end{document}